\documentstyle[12pt]{article}

\catcode`\@=11
\long\def\@makefntext#1{
\protect\noindent \hbox to 3.2pt {\hskip-.9pt
$^{{\ninerm\@thefnmark}}$\hfil}#1\hfill}		

\def\@makefnmark{\hbox to 0pt{$^{\@thefnmark}$\hss}}  

\def\ps@myheadings{\let\@mkboth\@gobbletwo
\def\@oddhead{\hbox{}
\rightmark\hfil\ninerm\thepage}
\def\@oddfoot{}\def\@evenhead{\ninerm\thepage\hfil
\leftmark\hbox{}}\def\@evenfoot{}
\def\sectionmark##1{}\def\subsectionmark##1{}}

\renewcommand{\thefootnote}{\fnsymbol{footnote}}

\def\refstepII#1#2{\stepcounter{#2}\let\@tempa\protect
\def\protect{\noexpand\protect\noexpand}%
\edef\@currentlabel{\csname p@#1\endcsname\csname the#1\endcsname%
.\csname p@#2\endcsname\csname the#2\endcsname}%
\let\protect\@tempa}
\def\refstepIII#1#2#3{\stepcounter{#3}\let\@tempa\protect
\def\protect{\noexpand\protect\noexpand}%
\edef\@currentlabel{\csname p@#1\endcsname\csname the#1\endcsname%
.\csname p@#2\endcsname\csname the#2\endcsname%
.\csname p@#3\endcsname\csname the#2\endcsname}%
\let\protect\@tempa}

\newcounter{sectionc}\newcounter{subsectionc}\newcounter{subsubsectionc}
\renewcommand{\section}[1] {\vspace*{0.6cm}\refstepcounter{sectionc}
\setcounter{subsectionc}{0}\setcounter{subsubsectionc}{0}\noindent
	{\normalsize\bf\thesectionc. #1}\par\vspace*{0.4cm}}
\renewcommand{\subsection}[1] {\vspace*{0.6cm}\refstepII{sectionc}{subsectionc}
	\setcounter{subsubsectionc}{0}\noindent
	{\normalsize\it\thesectionc.\thesubsectionc. #1}\par\vspace*{0.4cm}}
\renewcommand{\subsubsection}[1]
{\vspace*{0.6cm}\refstepIII{sectionc}{subsectionc}{subsubsectionc}
	\noindent {\normalsize\rm\thesectionc.\thesubsectionc.\thesubsubsectionc.
	#1}\par\vspace*{0.4cm}}

\newcounter{appendixc}
\newcounter{subappendixc}[appendixc]
\newcounter{subsubappendixc}[subappendixc]

\renewcommand{\appendix}[1] {\vspace*{0.6cm}
        \refstepcounter{appendixc}
        \setcounter{figure}{0}
        \setcounter{table}{0}
        \setcounter{equation}{0}
        \renewcommand{\thefigure}{\Alph{appendixc}.\arabic{figure}}
        \renewcommand{\thetable}{\Alph{appendixc}.\arabic{table}}
        \renewcommand{\theappendixc}{\Alph{appendixc}}
        \renewcommand{\theequation}{\Alph{appendixc}.\arabic{equation}}
        \noindent{\bf Appendix \theappendixc #1}\par\vspace*{0.4cm}}

\def\abstracts#1{{

\centering{\begin{minipage}{12.2truecm}\footnotesize\baselineskip=12pt\noindent
	\centerline{\footnotesize ABSTRACT}\vspace*{0.3cm}
	\parindent=0pt #1
	\end{minipage}}\par}}

\renewenvironment{thebibliography}[1]
	{\begin{list}{\arabic{enumi}.}
	{\usecounter{enumi}\setlength{\parsep}{0pt}
\setlength{\leftmargin 1.25cm}{\rightmargin 0pt}
	 \setlength{\itemsep}{0pt} \settowidth
	{\labelwidth}{#1.}\sloppy}}{\end{list}}

\newcounter{itemlistc}
\newcounter{romanlistc}
\newcounter{alphlistc}
\newcounter{arabiclistc}

\newcommand{\fcaption}[1]{
        \refstepcounter{figure}
        \setbox\@tempboxa = \hbox{\footnotesize Fig.~\thefigure. #1}
        \ifdim \wd\@tempboxa > 6in
           {\begin{center}
        \parbox{6in}{\footnotesize\baselineskip=12pt Fig.~\thefigure. #1}
            \end{center}}
        \else
             {\begin{center}
             {\footnotesize Fig.~\thefigure. #1}
              \end{center}}
        \fi}

\newcommand{\tcaption}[1]{
        \refstepcounter{table}
        \setbox\@tempboxa = \hbox{\footnotesize Table~\thetable. #1}
        \ifdim \wd\@tempboxa > 6in
           {\begin{center}
        \parbox{6in}{\footnotesize\baselineskip=12pt Table~\thetable. #1}
            \end{center}}
        \else
             {\begin{center}
             {\footnotesize Table~\thetable. #1}
              \end{center}}
        \fi}

\def\@citex[#1]#2{\if@filesw\immediate\write\@auxout
	{\string\citation{#2}}\fi
\def\@citea{}\@cite{\@for\@citeb:=#2\do
	{\@citea\def\@citea{,}\@ifundefined
	{b@\@citeb}{{\bf ?}\@warning
	{Citation `\@citeb' on page \thepage \space undefined}}
	{\csname b@\@citeb\endcsname}}}{#1}}

\newif\if@cghi
\def\cite{\@cghitrue\@ifnextchar [{\@tempswatrue
	\@citex}{\@tempswafalse\@citex[]}}
\def\citelow{\@cghifalse\@ifnextchar [{\@tempswatrue
	\@citex}{\@tempswafalse\@citex[]}}
\def\@cite#1#2{{$\null^{#1}$\if@tempswa\typeout
	{IJCGA warning: optional citation argument
	ignored: `#2'} \fi}}

 1
 1
 1

\font\ninerm=cmr9

\input psfig

\begin{document}
\quad\vskip-1.5cm
\rightline{\normalsize HEPSY-95-05}
\rightline{\normalsize November 1995}
\centerline{\normalsize\bf DECAYS OF {\Large $b$} QUARK}
\baselineskip=22pt
\centerline{\footnotesize TOMASZ SKWARNICKI}
\baselineskip=13pt
\centerline{\footnotesize\it Department of Physics,
Syracuse University}
\baselineskip=12pt
\centerline{\footnotesize\it Syracuse, NY 13244, USA}
\centerline{\footnotesize E-mail: tomasz@uhep2.phy.syr.edu}
\vspace*{0.3cm}
\abstracts{Importance of studies of $b$ quark decays and
           experimental status of various measurements
           are discussed.}
\vspace*{0.3cm}
\centerline{\footnotesize\it To appear in the Proceedings of}
\centerline{\footnotesize\it the $17^{th}$ International Conference on
            Lepton-Photon Interactions}
\centerline{\footnotesize\it Beijing, China, August 1995}
\normalsize\baselineskip=15pt
\setcounter{footnote}{0}
\renewcommand{\thefootnote}{\alph{footnote}}

\def\Vtb{V_{tb}}
\def\Vcb{V_{cb}}
\def\Vub{V_{ub}}
\def\Vts{V_{ts}}
\def\Vtd{V_{td}}
\def\Vcs{V_{cs}}
\def\Vus{V_{us}}

\def\BR{Br}

\def\lem{l^{^-}}
\def\lep{l^{^+}}

\def\ufs{\Upsilon(4S)}

\def\lowcite#1{\lower6pt\hbox{\Large\cite{#1}}}

\section{Introduction}

\subsection{Physics Goals}

The existence of three generations of quark weak doublets
has been well established experimentally.
The origins, however,  of different quark (and lepton) generations,
of their mixing in charged weak currents,
and of their mass spectrum have not been understood.
Furthermore, some elements of the
mixing matrix (the Cabibbo-Kobayashi-Maskawa matrix)
involving quarks of the
third generation ($b$ and $t$)
are poorly measured.
Decays of the $b$ quark
play a strategic role in
the determination of the CKM parameters.
The latter is very important for understanding the nature of
CP violation discovered in decays of $s$ quark, and for tests of
the CKM matrix unitarity (is there a fourth generation ?).
After the recent discovery of the top quark, the $b$ quark is no longer
the heaviest known quark, but it is still the heaviest among those
creating hadrons. As such, it plays a very important role in studies
of strong interactions.
Its heavy mass and relatively long lifetime, make decays of the $b$ quark
also an interesting place to search for rare decays due to
non-standard interactions.
In this paper we will concentrate on
recent experimental results
concerning rare decays and
determination of the CKM parameters.
The latter necessarily overlaps with understanding strong interaction
phenomena interfering with weak decays.

\subsection{Experimental Environments}

\begin{table}[bth]
\tcaption{Various parameters characterizing experimental environments
          at three colliders active in producing $b$ quarks.
          Explanation of the symbols: $\sigma-$cross section,
          ${\cal L}-$luminosity, $\beta-$velocity of $b$ quarks,
          $\beta\gamma c\tau-$mean decay path,
          $f-$fractions of $b$ hadron species produced.
}
\label{tab:benvir}
\small
\def\1#1{{#1}}
\def\2#1{{#1}}
\begin{tabular}{||l||r|r|r||}
\hline\hline
 {} & {} & {} & {} \\
Quantity      &  $\ufs$ (CESR)  & $Z^0$ (LEP) &   Tevatron \\
 {} & {} & {} & {} \\
\hline
 {} & {} & {} & {} \\
$\sigma(b\bar{b})$ nb$^{-1}$
  &   $1.1$  &   $9.2$  &  \1{$\sim30000$} $|y|<1$ \\
  &          &              &  (\1{$\sim6000$} $p_t\!>\!6$ GeV) \\
$\sigma(b\bar{b})/\sigma(q\bar{q})$ \%
           &  $\sim30$ & $\sim20$ & \2{$\sim0.1$} \\
${\cal L}^{peak}$ cm$^{-2}$ s$^{-1}$ 10$^{-31}$
                              &  \1{$33.0$}  &   \2{$1.1$}  &   $2.5$  \\
$\int{\cal L}dt$  pb$^{-1}$
               analyzed   &  \1{$2400$}  &  $\sim75$ &  $\sim65$ \\
\phantom{$\int{\cal L}dt$  pb$^{-1}$}
   in~pipeline              &  $1000$  &  $>13$    &  $>20$ \\
$b\bar{b}$ pairs   $10^6$   analyzed
                    &  $2.6$   &  $0.7$  &
$\sim2000$ $(\sim200)^{\dag}$
 \\
$\beta$ & \1{$\sim0.07$}   & $\sim1$ & $\sim1$ \\
$\beta\gamma c\tau$ $\mu m$ &  \2{30}   & \1{2600} & \1{500} \\
fragmentation & & & \\
\qquad\quad background   &  \1{no}  & some
  & \2{large} \\
       &  \1{$E_B=E_{beam}$}  & $\overline{E_B}\sim0.7\,E_{beam}$
  &    \\
spatial separation & & & \\
\qquad\quad of $b$ and $\bar{b}$
                   &  \2{no}  &  \1{yes}  &     \\
$f_{B^+}\approx f_{B^0}$  &  0.5  &  $\sim0.4$ & $\sim0.4$ \\
$f_{B^0_s}$      &  --  &  $\sim0.1$ & $\sim0.1$ \\
$f_{\Lambda_b}$  &  --  &  $\sim0.1$ & $\sim0.1$ \\
{} & {} & {} & {} \\
\hline
{} & {} & {} & {} \\
main advantage & simple production, & vertexing,         & cross-section, \\
               & statistics         & one $b$ at a time  & vertexing  \\
{} & {} & {} & {} \\
\hline\hline
\end{tabular}
\newline
${\dag}$\quad {These numbers correspond to
the central region, $|y|<1$, with high transverse momentum of
$b$ quark, $p_t>6$ GeV (the number in parentheses).}
\end{table}

At present $b$ quark
decays are under investigation
with large statistics at three different
colliders: CESR producing $B\bar{B}$ pairs in decays of the
$\Upsilon(4S)$ resonance just above the $e^+e^-\to B\bar{B}$ threshold,
LEP producing $b\bar{b}$ pairs in $Z^0$ decays, and Tevatron producing
$b\bar{b}$ pairs in $p\bar{p}$ collisions.
Various production aspects at these machines are compared in Table
\ref{tab:benvir}.
The CLEO-II experiment at CESR enjoys the highest achieved collider
luminosity and a simple production mechanism.
Both, charged tracks and photons, are detected with good
efficiency and resolution.
Since no fragmentation particles are produced, energy of reconstructed
$B's$ can be constrained to the beam energy which provides for a powerful
reconstruction technique. On the other hand since two $B$ mesons are
produced almost at rest, detection of a detached secondary $B$ decay
vertex is not possible on event by event basis. Also, decay products
from the two $B$'s in the event populate the entire solid angle.
At LEP momentum given to $b$ hadrons is appreciable, thus the decay
products from the two $b$ hadrons produced are easily separated into
back-to-back hemispheres. A large decay path gives rise to many
important analysis techniques. Even though the $b\bar{b}$ cross section
is much larger than at the $\Upsilon(4S)$ resonance, a high energy
of the beam limits achievable luminosity. In view of the forthcoming
end of the $Z^0$ running at LEP, statistics should be considered the
main limiting factor for the LEP experiments.
The huge production cross-section
is the main asset of the Tevatron experiments.
A large background cross-section is the main obstacle to overcome in
$b$ experiments with hadronic beams.
Specialized triggering is needed to limit data acquisition to
manageable rates.
So far, the results from Tevatron
have been mostly limited to channels containing high $p_t$ muons.
Again, vertexing is an important selection tool.

\subsection{General structure of $b$ quark decays
in the Standard Model.}

\begin{figure}[htbp]
\quad\hskip1cm\psfig{file=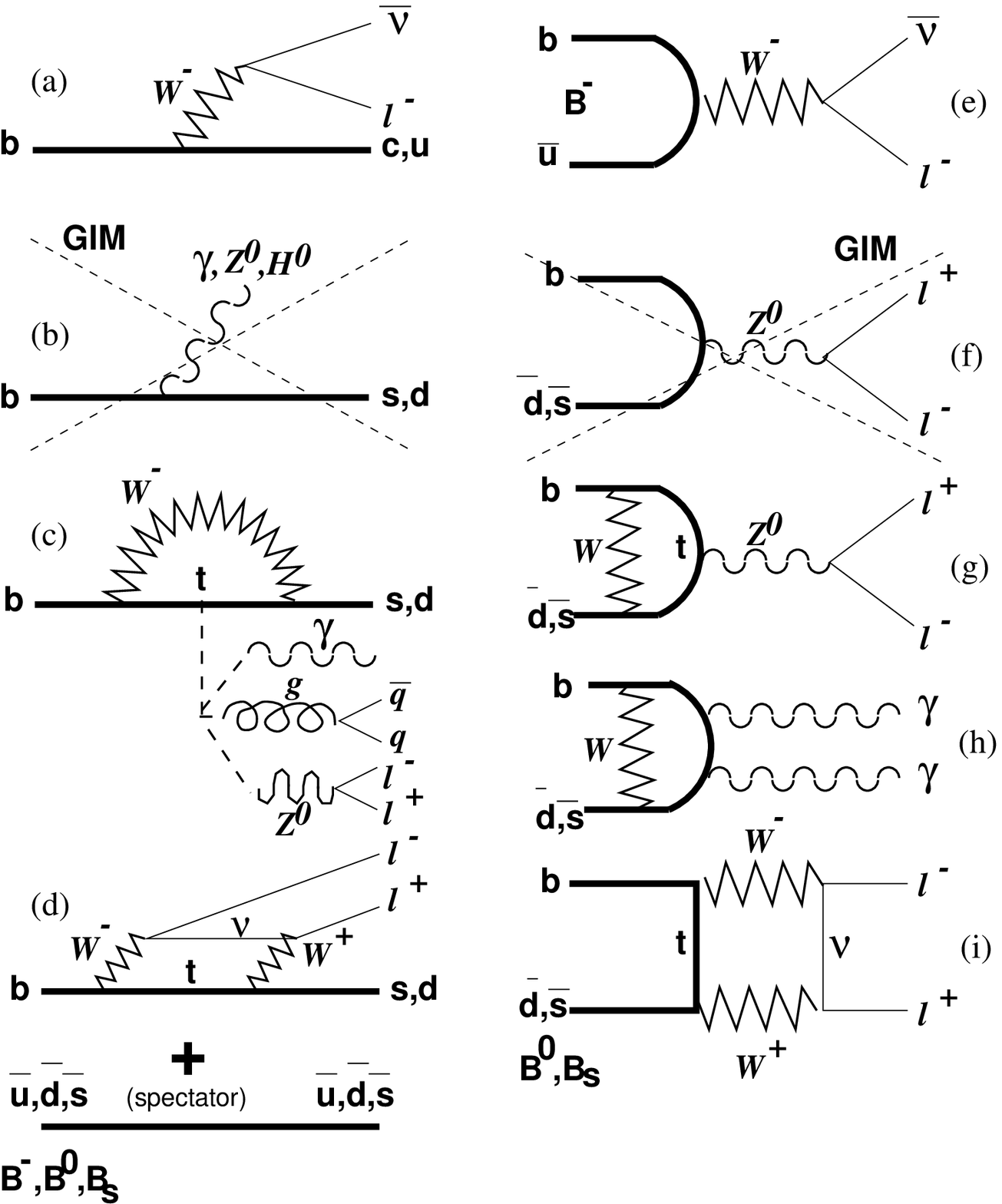}
\fcaption{Diagrams describing various decays of $B$ mesons.
          Diagrams (b) and (f) are forbidden by the GIM mechanism.
          Any lepton line can be also replaced by quark line.}
\label{fig:bdec}
\end{figure}

Since decays of the $b$ quark to the $t$ quark and $W^-$ are kinematically
forbidden in the first order, the $b$ quark decays to the $c$ or $u$
quark and virtual $W^-$ that turns to a pair of leptons or
quarks (Fig.~\ref{fig:bdec}a).
The decays to the $c$ quark are strongly favored by
the values of the CKM elements ($|\Vcb|^2/|\Vub|^2\sim10^2$).
Decays to down-type quarks, $s$ or $d$,
are forbidden by the GIM mechanism (Fig.~\ref{fig:bdec}b).
In the second order, $b$ can decay to virtual $t$ and $W^-$ that
immediately recombine to the $s$ or $d$ quark (so called loop or penguin
decays).
The decays to the $s$ quark
are favored by the CKM elements ($|\Vts|^2/|\Vtd|^2\sim10^1$).
To conserve energy, something
must be emitted from these interactions: a photon, a gluon that
turns to quarks, or a neutral $Z^0$ that turns to a pair of quarks, leptons
or neutrinos (Fig.~\ref{fig:bdec}c).
 The latter decays can also occur via a box diagram
with two charged $W$'s exchanged between the heavy and the lighter fermion
lines (Fig.~\ref{fig:bdec}d).

All of the above are spectator decays, since the other quark
in the initial hadron does not participate in the short distance disintegration
of the $b$ quark. Analogous decays exist for non-spectator decays of
$B_q$ mesons in which the $b$ quark and its lighter partner $q$ annihilate
to virtual bosons (Fig.~\ref{fig:bdec}e-i).
Annihilation processes are suppressed
by the smallness of the $B$ decay constant.
Furthermore, since the $b$ quark is heavy,
helicity suppression is very effective
in annihilation to leptons.
In the first order, only $B^-_u$ can annihilate (Fig.~\ref{fig:bdec}e).
However, this process is
strongly disfavored by the smallness of the CKM element involved ($\Vub$).
Annihilation of $B^0_d$ and $B_s$ can occur in the second order by the
inverted penguin or box diagrams (Fig.~\ref{fig:bdec}g-i).
The annihilation of $B_s$
is preferred by the CKM element ($|\Vts|^2/|\Vtd|^2\sim10^1$).
As a combined effect of the suppression factors
mentioned above, the annihilation processes
in $B$ meson decays are expected to be very rare.
Therefore, the spectator decays of the $b$ quark are easier to detect.
The first order spectator decays serve a measurement
 of the third column of the CKM
matrix ($|\Vcb|$ and $|\Vub|$), whereas the loop decays can be used to probe
the third raw of the matrix ($|\Vts|$ and $|\Vtd|$).

An alternative way to determine $|\Vtd|$ and $|\Vts|$ from $b$
interactions is to
study $B^0\bar{B}^0$ and $B_s\bar{B}_s$ oscillations, produced by the box
diagrams. Complex phases of the CKM elements can be accessed via CP violating
phenomena expected in $b$ quark decays. The above topics are the subject of
separate papers presented at this conference.~\cite{other}


\begin{figure}[thbp]
\psfig{file=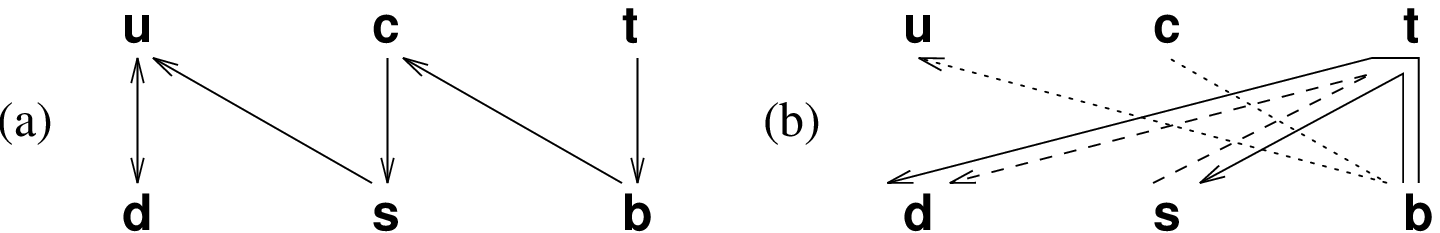}
\fcaption{Standard Model decays of quarks. (a) Dominant first
          order decays. (b) Second order quark transitions.}
\label{fig:qvq}
\end{figure}

\subsection{Decays of the $b$ quark versus decays of the other quarks}

Decays of down-type quarks are generally more interesting than decays
of up-type quarks.
Because of the quark mass pattern, up-type quarks can decay to a down
type partner within the same generation (Fig.~\ref{fig:qvq}a).
Thus, dominant decays of these
quarks probe diagonal CKM elements that are known to be close to unity.
These decays make the up-type quarks relatively short lived.
In contrast, dominant decays of heavy down-type quarks change the
quark generation. Because of the smallness of the off-diagonal CKM elements,
the down-type quarks are relatively long lived.
Decays of up-type quarks to the quarks of the other generations are rare and
difficult to observe on top of the background from the dominant decays.
Decays of the $t$ quark to the $s$ or $d$ quark will be particularly
difficult to observe.
At present,
the decays of the $b$ and $s$ quarks are
the best experimental path
to measure $|\Vts|$ and $|\Vtd|$.

The down-type quarks are also favored over the
up-type quarks in the Standard Model
loop decays (Fig.~\ref{fig:qvq}b).
Since rate increases with the mass of the particle in
the loop, exchanges of the $t$ and $b$ quarks dominate.
Top as the heaviest quark has nothing to exchange in a virtual loop.
The charm quark may
decay to the $u$ quark by exchanging the $b$ quark, but rate for
this decay is predicted to be extremely small since the $c$ quark is short
lived and the CKM factor ($|\Vcb\cdot\Vub^*|^2$) is tiny.
This leaves the
$b$ quark decays the only way to measure $|\Vcb|$ and $|\Vub|$.

The strange quark can decay to the $d$ quark by exchanging the $t$ quark.
As a down-type quark the $s$ quark is longer lived.
Also the CKM suppression
is expected to be somewhat smaller ($|\Vtd|>|\Vub|$).
Thus, the loop decays of the $s$ quark are an interesting way to probe
$\Vts\cdot\Vtd^*$.

Even a better access to these elements is provided by the
loop decays of the $b$ quark.
There are two possible loop decays:
to the $s$ quark and to the $d$ quark, thus each element
can be probed separately.
Since there is no CKM suppression in the
intermediate $b\to t$ transition, the loop
decays of the $b$ quark are favored by very large factors compared
 to the $s$ decays:
$(b\to s)/(s\to d)\sim|(\Vtb\cdot\Vts^*)/(\Vts\cdot\Vtd^*)|^2
\sim1/|\Vtd|^2\sim10^4$,
$(b\to d)/(s\to d)\sim  1/|\Vts|^2\sim10^3$ ).
Also, the larger suppression of the dominant decays favors the $b$ quark
( $|\Vus|^2/|\Vcb|^2\sim30$ ).
The large CKM suppression factors can be overcome in $s$ decay experiments
by a higher quark production rate and a more controllable environment than
in $b$ decay measurements.
However, the large CKM suppression is to be blamed
for long distance strong interactions overwhelming almost all
rare kaon decay channels.\footnote{
The only exceptions are: $K^+\to\pi^+\nu\bar{\nu}$
($BR\sim10^{-10}$), $K^0_L\to\pi^0\nu\bar{\nu}$
($BR\sim10^{-11}$), and any CP violating pieces of rare or usual kaon decays.}
For example, the $K^0_L\to\gamma\pi\pi$ decay, that can proceed via
an electromagnetic
penguin decay $s\to d\gamma$, is completely dominated by an internal
$W$ exchange
producing a virtual $\pi^0$ or $\eta$
that decays to $\gamma\pi\pi$.~\cite{rarek}
In contrast, long distance strong interactions contribute only about 4\%\ to
channels produced by generic $b\to s\gamma$ decays.~\cite{bsgLD}
The short distance interactions are also expected to dominate
$b\to d$ channels, though long distance contributions may be larger here.

Similarly, the loop decays of the $b$ quark offer better sensitivity
than the $s$ quark decays for
various contributions beyond the Standard Model,
especially if mass dependent couplings are involved.

In conclusion, decays of the $b$ quark are the most interesting among all
decays of quarks.

\section{Rare decays of $b$ quark}

\subsection{Electromagnetic penguin decays}

\begin{figure}[htbp]
\vbox{
\quad\vskip-1.5cm
\hbox{(a) \hskip-1cm
\lower7.5cm\psfig{file=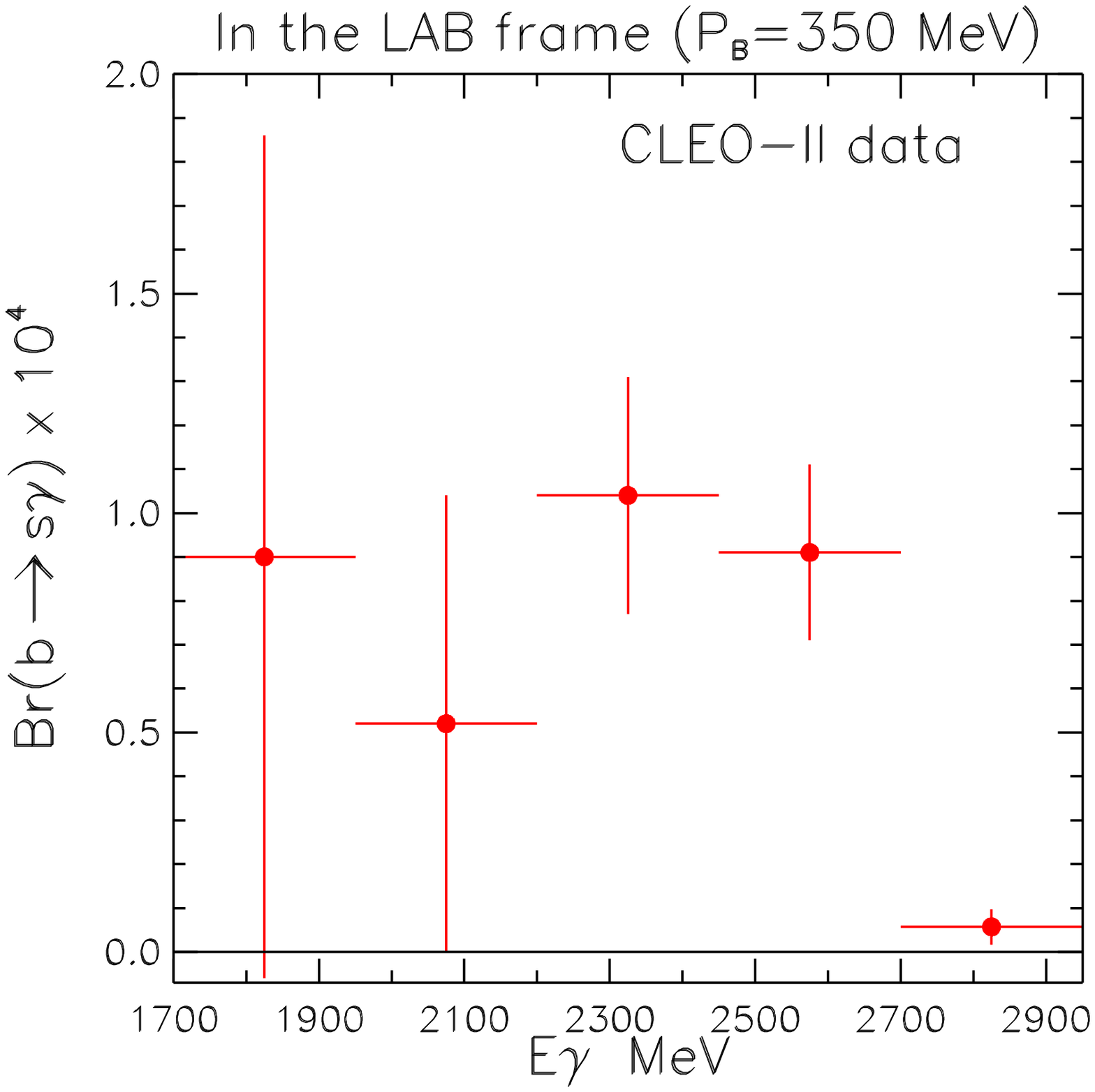,width=9cm,height=9cm}\hskip-1cm
(b) \hskip-1cm
\lower7.5cm\psfig{file=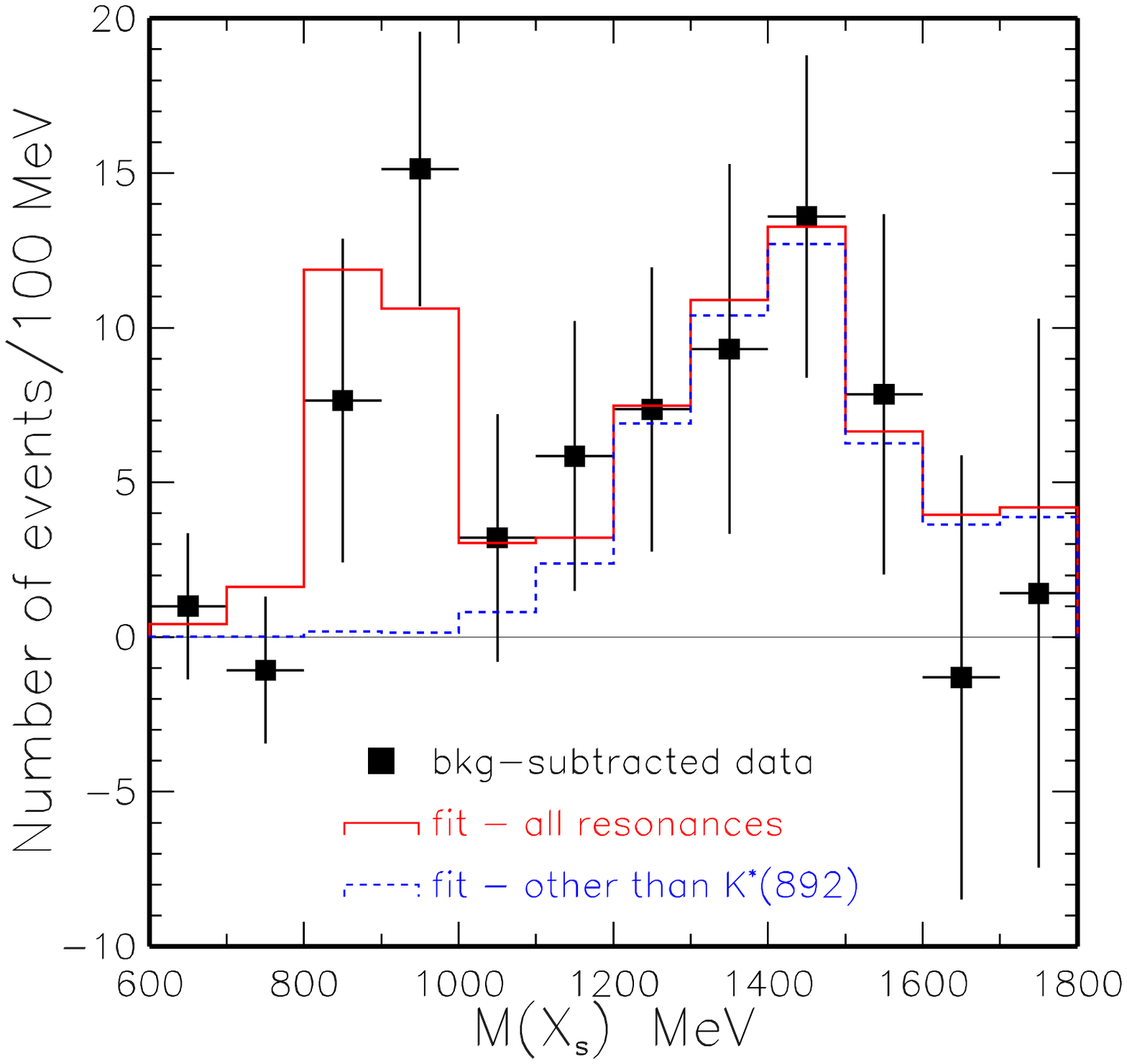,width=9cm,height=9cm}
}
\vskip-1.5cm
\quad}
\fcaption{Inclusive $b\to s\gamma$ signal in the CLEO-II data.
(a) Measured photon spectrum averaged over the two analyses
    presented by CLEO.~\cite{CLEObsg}
(b) Mass of hadronic system recoiling the photon.}
\label{fig:CLEObsg}
\end{figure}

\subsubsection{$b\to s\gamma$}
\label{sec:bsg}

Electromagnetic penguin decays $b\to s\gamma$ were first observed by
the CLEO-II experiment in the exclusive decay
mode $B\to K^*\gamma$ in a sample of $1.5\times10^6$ $B\bar{B}$
pairs.~\cite{CLEOKsg93}
Since then, CLEO-II has almost doubled the sample and
presents
an updated branching fraction for $B^0\to K^{*0}\gamma$.~\cite{CLEOKsgEPS}\
Combining this result with the original measurement of
$B^-\to K^{*-}\gamma$, we obtain:
$\BR(B\to K^{*}\gamma)=(4.5\pm1.0\pm0.6)\times10^{-5}$.

CLEO-II recently published an inclusive measurement of the total
$b\to s\gamma$ rate.~\cite{CLEObsg}
Using the data sample of $2.2\times10^6$ $B\bar{B}$ pairs
CLEO-II measured: $\BR(b\to s\gamma)=(2.32\pm0.57\pm0.35)\times10^{-4}$.
With a quarter of that statistics, DELPHI sets an
upper limit on this branching ratio, $<6.3\times10^{-4}$ (90\%\ C.L.),
which is consistent with the CLEO-II value.~\cite{DELPHIbsgEPS}
The inclusive photon spectrum
averaged over the two
selection methods used by CLEO is displayed in
Fig.~\ref{fig:CLEObsg}a.
As expected for quasi two-body decay, the spectrum peaks
at $m_b/2$.
Exact predictions for the photon energy distribution in this
process are a subject of recent theoretical papers.~\cite{AliG95eg,bsgeg}

As illustrated in Fig.~\ref{fig:CLEObsg}b,
the CLEO-II inclusive measurement clearly indicates
that the $b\to s\gamma$ process produces final states with
masses also above the $K^*$ resonance, though with the present
experimental statistics it is not possible to disentangle
various resonances in the higher mass region.
The ratio of branching ratios,
$R_{K^*}=\BR(B\to K^*\gamma)/\BR(b\to s\gamma)=(19\pm6\pm4)\,\%$,
depends on the long distance strong interactions that follow
the short distance penguin decay and create the $K^*$ resonance.
Additional complication arises from a possibility that
the photon in this decay mode could also be generated by
long distance strong interactions rather than the penguin
decay (e.g. via formation of virtual $J/\psi$ or $\rho$
turning to the photon).
The $R_{K^*}$ ratio
has been predicted by a number of phenomenological models and
by QCD on lattice.
Recent predictions are in rough agreement with the data.~\cite{ksksg}

More interesting is the inclusively measured branching fraction
since it does not depend on the hadronization process, and therefore
reflects short distance interactions.\footnote{
See Sec.\ref{sec:isbc}\ for more detailed discussion of
advantages of inclusively measured rates.}
A value of $|\Vts|$ can be extracted from a ratio of the
the measured  branching ratio and the Standard Model prediction:
$|\Vts|^2/|\Vcb|^2=\BR(b\to s\gamma)_{meas.}/
\BR(b\to s\gamma)_{SM}$.\footnote{$\BR(b\to s\gamma)_{SM}$ denotes
here the Standard Model calculation that assumes $|\Vts|^2/|\Vcb|^2=1$.}
Using the calculations by Buras et al.~\cite{Buras}\ we
obtain:
$$
\frac{|\Vts|}{|\Vcb|} = 0.91\pm0.12(experimental)\pm0.13(theoretical)
$$
in agreement with $\sim1.0$ predicted by assuming unitarity of the
CKM matrix.
Larger estimates of the
theoretical uncertainties can be found.~\cite{AliG95eg}
The main theoretical error is due to the renormalization scale
dependence since no complete next-to-leading order calculation
exists. The theoretical error is already larger than the
experimental one, thus complete next-to-leading order
calculations would be
very welcome.

Assuming the CKM matrix unitarity, the above ratio can be turned
into an upper limit on the interactions beyond the Standard Model.
Since the Standard Model diagram exchanges the two heaviest objects
we know so far, the top quark and the $W$ boson, the $b\to s\gamma$ process
is very sensitive to a possible exchange of the non-standard particles,
if such exist on a mass scale not too much larger than the scale of
the electroweak interactions.
A large number of extensions of the Standard Model have been discussed
in this context.~\cite{Hewett}
A typical example is a Two-Higgs-Doublet-Model in which the $W$  in the
penguin loop can be
replaced by a charged Higgs.
Since the Higgs contribution always increases the predicted rate,
a lower limit on the Higgs mass
can be obtained:
$M_{H^-}>[244+63/(\tan\beta)^{1.3}]$ GeV, 
where $\tan\beta$ is a ratio of the vacuum expectations for the
two Higgs doublets.\footnote{This limit is valid for type-II
Two-Higgs-Doublet-Models. See e.g.\ Ref~\lowcite{Hewett}\
for more detailed discussion.}
For theoretically preferred large values of $\tan\beta$ the limit is
almost constant.
In supersymmetric models chargino and stop
can be exchanged as well,
that may add constructively or destructively to the total rate.
Thus limits imposed on the supersymmetric parameters are somewhat
complex. Naively, the charged Higgs, chargino and stop are either all
heavy or all light.

\subsubsection{$b\to d\gamma$}

The main experimental difficulty in the search for $b\to d\gamma$
transitions is to distinguish them from more frequent
$b\to s\gamma$ decays.
CLEO-II searched for exclusive $B\to(\rho/\omega)\gamma$
decays.~\cite{CLEOdsg}
Because of the simplicity of the final state, the $B\to K^*\gamma$
background could be efficiently suppressed by kinematical cuts.
With the present statistics no signal was found and an upper limit
on $\BR(B\to(\rho/\omega)\gamma)/\BR(B\to K^*\gamma)$ was set.
Neglecting possible long distance interactions, the latter ratio is
equal to $\xi |\Vtd|^2/|\Vts|^2$, where $\xi$ is a
model dependent factor due to
the phase space and SU(3)$_f$ breaking.~\cite{alirhog}
Depending on the value of $\xi$:
$$ \frac{|\Vtd|}{|\Vts|}<0.64-0.76 $$

It has been pointed out by many authors that long distance
contributions may be significant for these decays.~\cite{bdgLD}
Separate measurements of $B^-$ and $B^0$ decays, and of $\rho\gamma$
and $\omega\gamma$ will be helpful in isolating short and long
distance contributions.

\subsection{Rare exclusive hadronic decays}

Gluonic penguin decays should give rise to various rare hadronic
decay modes of the $b$ quark. The total inclusive rate for such decays is
expected to be much larger than for the electromagnetic penguin
decays. Nevertheless these decays are more difficult to observe
experimentally, since there is no common inclusive signature for these
decays and exclusive branching ratios are small due to a large number
of possible decay modes. Furthermore, many of these decay modes can be
produced also by the $b\to u W^-$ decays which adds complexity to
the analysis.

The main interest in rare hadronic decay modes is their future application
in the determination of phases of the CKM matrix elements via measurements
of the CP asymmetries.~\cite{StoneP}\
Interference of penguin and $b\to u$ amplitudes
should give rise to direct CP violation in decays of charged and
neutral $B$ mesons. Interference of unmixed and mixed decays of $B^0$
to a CP eigenstate, like e.g. $B^0\to\pi^+\pi^-$, is expected to
produce indirect CP violation.
\begin{figure}[htbp]
\vbox{
\quad\vskip-2.5cm
\hbox{\hskip-2.5cm
\lower12.0cm\psfig{file=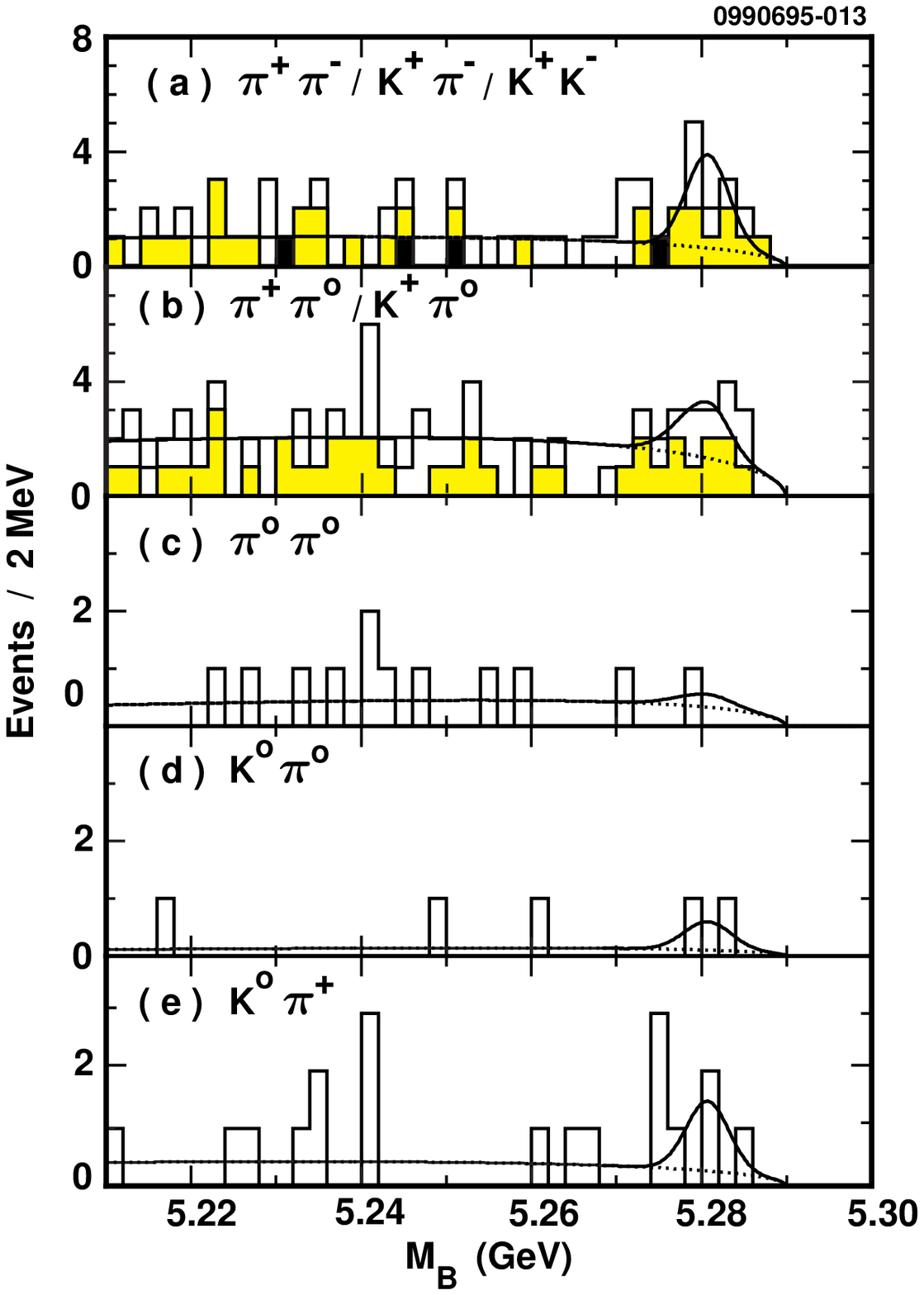,width=13cm}\hskip-3.0cm
(f) \hskip-0.5cm
\lower7.5cm\psfig{file=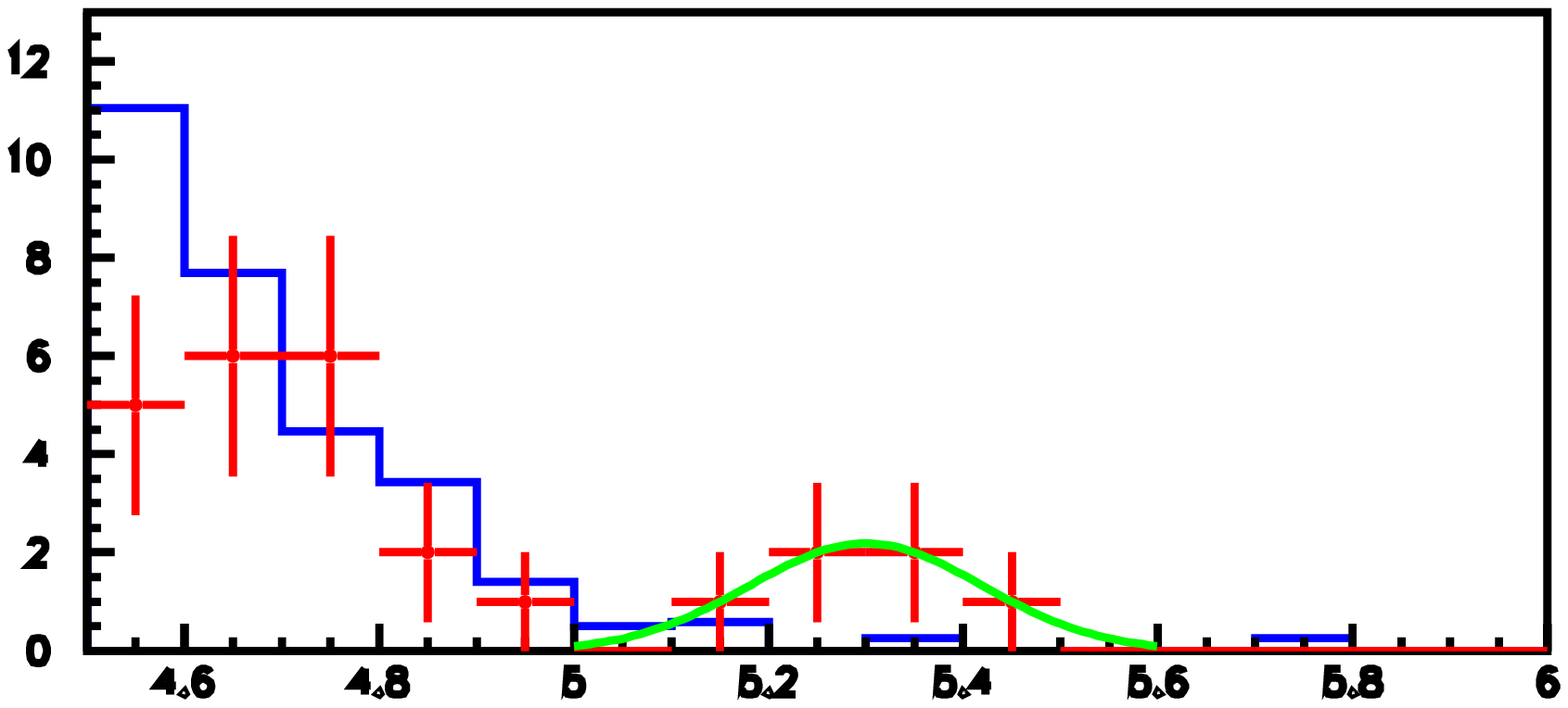,width=8cm}
}
\vskip-4.5cm
\quad}
\fcaption{Invariant mass of candidates for $B$ decay products in
searches for charmless hadronic decays.
(a-e) A sample of CLEO-II results.
      Subsamples with preferred $K\pi$/$\pi\pi$/$KK$ interpretation
      are indicated by grey/unshaded/black histograms respectively.
(f) Search for two-body and quasi two-body decay modes from DELPHI.
Candidates for $\pi^+\pi^-$/$K^+\pi^-$/$K^+K^-$ /
$\rho^0\pi^-$/$K^{*0}\pi^-$/$K^-\rho^0$/$K^-\phi$/$K^+a_1^-$
are plotted together.
The points with error bars represent
the data and the histogram represents the estimated background
from the $b\to c$ decays.
Note that the DELPHI plot covers a 15 times larger
range in $M_B$ than the CLEO-II pictures.
Neutral decay modes can be produced only by $B^0$ decays in the
CLEO-II data, whereas at LEP they can come from
either $B^0$ or $B_s$ decays.}
\label{fig:CLEObigrare}
\label{fig:DELPHIrare}
\end{figure}

Three years ago CLEO-II reported~\cite{CLEOhh93}\
the first evidence for a significant
signal in $B^0$ decays to $h^+\pi^-$, where $h^+$ denotes either
$\pi^+$ or $K^-$.
An updated measurement~\cite{CLEObigrare}\
with doubled the data sample ($2.6\times10^6$
$B\bar{B}$ pairs)  is
$\BR(B^0\to h^+\pi^-)=(1.8{^{+0.6}_{-0.5}}{^{+0.3}_{-0.4}})\times10^{-5}$.
About 17 signal events are observed.
The most recent estimate of the $\pi^+\pi^-$ component of these decays is
$\BR(B^0\to \pi^+\pi^-)/\BR(B^0\to h^+\pi^-)=(54\pm20\pm5)\%$.
Since neither the $\pi^+\pi^-$ nor the $K^+\pi^-$ channel reaches three
standard deviations in signal significance, CLEO-II quotes upper
limits for these modes taken separately.
There is an accumulation of events at the $B$ mass for some
other two-body and quasi two-body decay
modes as shown in Fig.~\ref{fig:CLEObigrare}a-e, though
none of these peaks is statistically significant, thus
upper limits are presented.

The LEP experiments have also searched for rare hadronic decay modes of $B$
mesons. For example, DELPHI~\cite{DELPHIbsgEPS}\
in a sample of $0.7\times10^6$ $b\bar{b}$
pairs produced observes a signal of 6 events over the estimated
background of 0.3 events when all two-body and quasi two-body decays are
put together.
Half of these candidates are due to $h^+\pi^-$ events
which is consistent with the rate measured by CLEO-II.
As illustrated in Fig.~\ref{fig:DELPHIrare}f the $B$ mass peak
is much broader at LEP since $B$ energy cannot be constrained to the
beam energy as in the CLEO-II experiment. On the other hand backgrounds
at LEP are smaller thanks to vertex cuts and separation of decay
products from the two $B$ mesons produced.

\begin{figure}[htbp]
\vbox{
\quad\vskip-6cm
\hbox{\quad\hskip-1.0cm\psfig{file=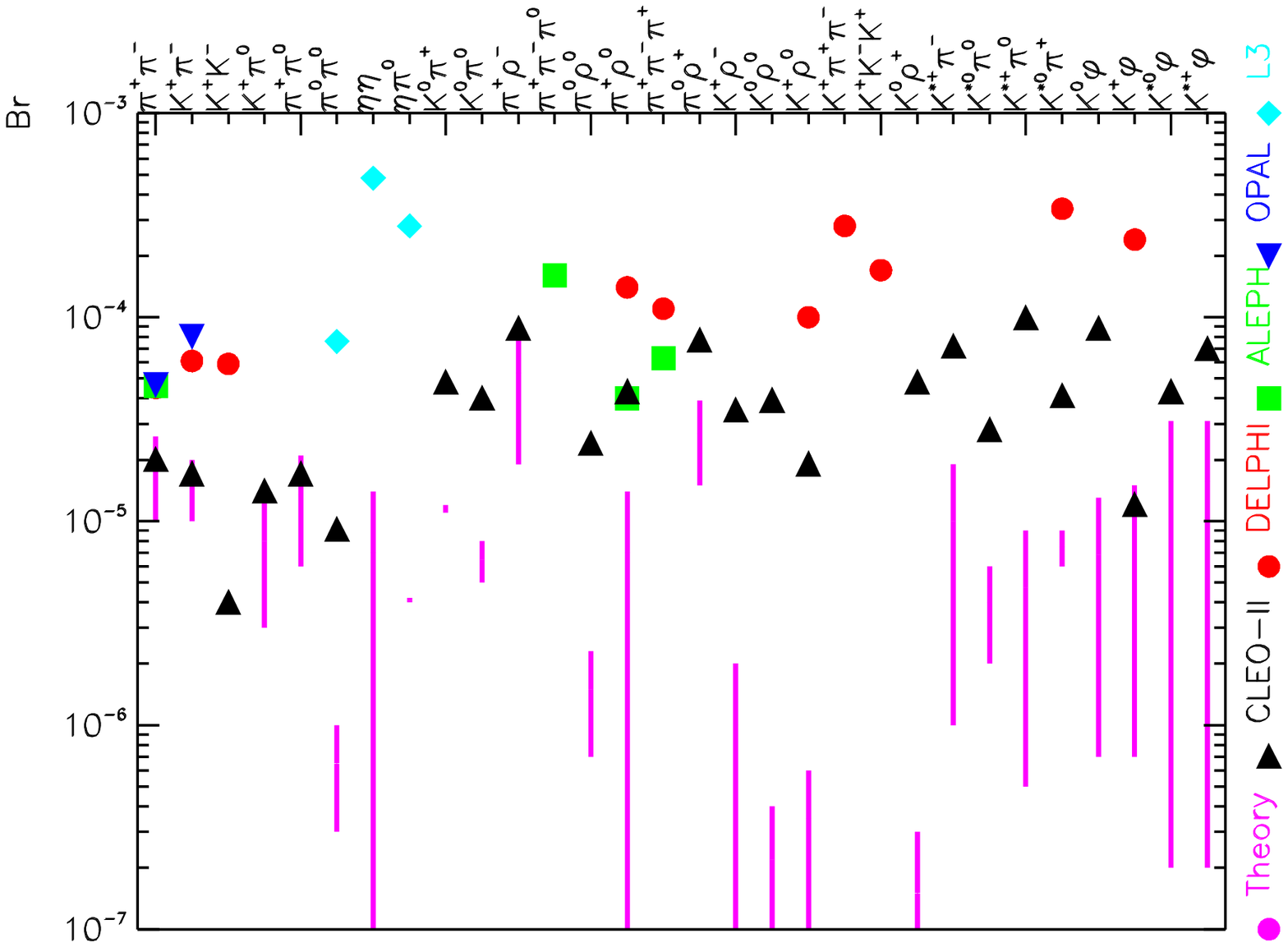,width=17cm}}
\vskip-7cm
\quad}
\fcaption{Upper limits (at 90\%\ C.L.)
on branching fractions for
charmless hadronic decays of $B$ mesons obtained by various
experiments. Vertical lines show a range of theoretical
predictions when available. The predictions are model dependent.
The CLEO-II analysis (triangles)
is based on $2.6\times10^6$ $B\bar{B}$ pairs.
The ALEPH (squares), DELPHI (circles) and L3 (diamonds)
data samples included $0.7\times10^6$
$b\bar{b}$ events. The OPAL results (inverted triangles)
are based on $0.4\times10^6$
$b\bar{b}$ pairs produced.}
\vskip-2cm
\label{fig:rare}
\end{figure}

Upper limits reported by CLEO-II~\cite{CLEObigrare}, ALEPH~\cite{ALEPHrare},
DELPHI~\cite{DELPHIbsgEPS},
L3~\cite{L3rare}, and OPAL~\cite{OPALrare}\ for various decay modes
are illustrated in Fig.~\ref{fig:rare}.
Ranges of branching ratios predicted by theoretical models are also
indicated. Thanks to the highest $b$ quark statistics,
the limits obtained by CLEO-II are the most stringent. L3 extended the
searches to channels with $\eta$, and ALEPH and DELPHI to channels
with multiplicities of up to 6 particles.


\subsection{Search for inclusive $b\to s\phi$}

Gluonic penguin decays which produce three strange
quarks, $b\to sg$, $g\to s\bar{s}$ have no competition from
the other decay mechanisms.
Assuming that $\phi$ meson can be directly produced from these
short distance interactions, as advocated by
Deshpande and He,~\cite{phixsDesh}
we can talk about $b\to s\phi$ decays.
Such decays would have characteristics similar to those of
$b\to s\gamma$, in particular effective quasi two-body
kinematics leading to a hard $\phi$ momentum spectrum and a small
effective mass of the recoiling hadronic system ($X_s$).
Using the analysis method previously applied to measure inclusive
$B\to X_s\gamma$ decays, CLEO-II finds no evidence for
$B\to X_s\phi$ decays with the present data sample and sets an
upper limit on the branching ratio of $<1.1\times10^{-4}$ (90\%\ C.L.)
for $M(X_s)<2$ GeV. Using recent predictions
of Deshpande et al.~\cite{EphixsDesh}
the limit can be extrapolated to the total rate:
$\BR(b\to s\phi)<1.3\times10^{-4}$ (90\%\ C.L.).
This result is consistent with the theoretical predictions by
Deshpande and He,~\cite{phixsDesh}
and excludes half of the predicted range:
$(0.6-2.0)\times10^{-4}$.

\subsection{Search for electroweak penguin decays}

The process $b\to s \lep\lem$ ($l=e$ or $\mu$) can occur via
electroweak penguin or box diagrams.
Experiments searched for these decays in the exclusive modes
$B\to K \lep\lem$ and $B\to K^* \lep\lem$.
In addition to earlier UA1~\cite{UA1}\
and CLEO-II~\cite{CLEOKll}\ results,
there are updated results presented by CDF.~\cite{CDFKll}
The experiments at hadronic colliders are competitive in this channel
with CLEO-II because of high $b$ production rates and
manageable backgrounds in dimuon samples, especially after
identification of the $B$ decay vertex.
The decays
$B\to K^{(*)}J/\psi$, $J/\psi\to \lep\lem$
due to $b\to c W^-$, $W^-\to \bar{c} s$ and binding of the
$c\bar{c}$ pair to $J/\psi$
by long distance strong interactions
that have a well measured branching ratio, are used as
a normalization mode. The experimental upper limits are an order
of magnitude away from the
theoretical predictions (that include calculation of
$K$ and $K^*$ form factors)
as shown in Table \ref{tab:kll}.
\begin{table}[hbtp]
\tcaption{Searches for electroweak penguin decays and
          for annihilation of $B$ mesons.
          Upper limits are given at 90\%\ C.L.}
\label{tab:kll}
\small
\def\1#1{\multicolumn{1}{l|}{#1}}
\def\2#1{\multicolumn{2}{l||}{#1}}
\begin{tabular}{||l||l|l||l||l||l||}
\hline\hline
   & \multicolumn{2}{c||}{}
   & \multicolumn{2}{c}{}
   & {} \\
   & \multicolumn{2}{c||}{$Br(B^-\to K^-\lep\lem)$}
   & \multicolumn{2}{c}{$Br(B^0\to K^{*0}\lep\lem)$}
   & {} \\
\cline{2-6}
   &  $e^+e^-$  & $\mu^+\mu^-$ &  \1{$e^+e^-$}  & \2{$\mu^+\mu^-$} \\
  {} & {} & {} & \1{} & \2{} \\
\hline
  {} & {} & {} & \1{} & \2{} \\
S.M. &  $0.04-0.08$  & $0.04-0.08$
     & \1{$0.5-0.7$} & \2{$0.3-0.4$} \\
   &  $\quad\times10^{-5}$ & $\quad\times10^{-5}$
   &  \1{$\quad\times10^{-5}$} & \2{$\quad\times10^{-5}$}    \\
  {} & {} & {} & \1{} & \2{} \\
\hline
  {} & {} & {} & \1{} & \2{} \\
UA1\cite{UA1} & {} & {} & \1{} & \2{$<2.3\times10^{-5}$} \\
CLEO-II\cite{CLEOKll}
  & $<1.2\times10^{-5}$
  & $<0.9\times10^{-5}$
  & \1{$<1.6\times10^{-5}$}
  & \2{$<3.1\times10^{-5}$} \\
CDF\cite{CDFKll}
  & {}
  & $<1.1\times10^{-5}$
  & \1{}
  & \2{$<2.1\times10^{-5}$} \\
  {} & {} & {} & \1{} & \2{} \\
\hline\hline
 {} & \multicolumn{2}{c||}{}
    & \multicolumn{1}{c||}{}
    & \multicolumn{1}{c||}{}
    & \multicolumn{1}{c||}{} \\
 {}  & \multicolumn{2}{c||}{$\BR(B^0\to \lep\lem)$}
     & \multicolumn{1}{c||}{$\!\BR(B^0_s\to \lep\lem)\!$}
   & \multicolumn{1}{c||}{$\!Br(B^0\to\gamma\gamma)\!$}
   & \multicolumn{1}{c||}{$\!Br(B^0_s\to\gamma\gamma)\!$} \\
\cline{2-4}
 {}  &  {$e^+e^-$} & {$\mu^+\mu^-$} & {$\mu^+\mu^-$} & {} & {} \\
 {}  &  {} & {} & {} & {} & {} \\
\hline
 {}  &  {} & {} & {} & {} & {} \\
S.M.  & $\phantom{ < }2\times10^{-15}$ &
        $\phantom{ < }8\times10^{-11}$ &
        $\phantom{ < }2\times10^{-9}$  &
        $\sim10^{-8}$              &
        $\sim10^{-7}$ \\
 {}  &  {} & {} & {} & {} & {} \\
\hline
 {}  &  {} & {} & {} & {} & {} \\
UA1\cite{UA1}        & {} & $<8.3\times10^{-6}$ & {}
   & {} & {} \\
CLEO-II\cite{CLEOll} & $<5.9\times10^{-6}$ & $<5.9\times10^{-6}$ & {}
   & {} & {} \\
CDF\cite{CDFll}      & {} & $<1.6\times10^{-6}$ & $<8.4\times10^{-6}$
   & {} & {} \\
L3\cite{L3gg}        & {} & {}              & {}
   & $<3.8\times10^{-5}$ & $<1.1\times10^{-4}$  \\
 {}  &  {} & {} & {} & {} & {} \\
\hline
\hline
 {}  & \multicolumn{3}{c||}{}
     & \multicolumn{2}{c||}{} \\
 {}  & \multicolumn{3}{c||}{$Br(B^-\to \lem\bar{\nu}_l)$}
     & \multicolumn{2}{c||}{$Br(B^-\to \lem\bar{\nu}_l\,\gamma)$} \\
\cline{2-6}
   &  $e^-$ & \1{$\mu^-$} &  $\tau^-$
   &  \1{$e^-$} & $\mu^-$ \\
 {} & {} & \1{} & {} & \1{} & {} \\
\hline
 {} & {} & \1{} & {} & \1{} & {} \\
S.M.  & $\sim10^{-13}$ & \1{$\sim10^{-8}$} &  $\sim10^{-5}$ &
        \1{$\sim10^{-6}$}  & $\sim10^{-6}$ \\
 {} & {} & \1{} & {} & \1{} & {} \\
\hline
 {} & {} & \1{} & {} & \1{} & {} \\
CLEO-II\cite{CLEOln,CLEOlng} &
  $<1.5\times10^{-5}$     &
  \1{$<2.1\times10^{-5}$}     &
  $<2.2\times10^{-3}$     &
  \1{$<1.6\times10^{-4}$} &
  $<1.0\times10^{-4}$ \\
ALEPH\cite{ALEPHln} &
   {} & \1{} &
   $<1.8\times10^{-3}$ &
   \1{}  & {} \\
 {} & {} & \1{} & {} & \1{} & {} \\
\hline\hline
\end{tabular}
\end{table}

\subsection{Search for $B$ meson annihilation}

Table \ref{tab:kll} also
summarizes results and predictions for the
annihilation processes.

CDF presents~\cite{CDFll}\
a search for $B^0$ and $B_s$ annihilation to
$\mu^+\mu^-$.
Even though the analysis is restricted to candidates in the central
region ($|y|<1$) with high transverse momenta
($p_t>6$ GeV) there are about $10^8$ $B^0$'s and $3\times10^7$ $B_s$'s
produced in this region in the CDF data sample.
CDF sets the first upper limit on $B_s\to\mu^+\mu^-$ branching
fraction and sets the limit on $B^0\to\mu^+\mu^-$ rate which is
a factor of three more stringent than the CLEO-II results~\cite{CLEOll}\
based on about $1.6\times10^6$ $B^0$'s.

The L3 experiment presents~\cite{L3gg}\
the first upper limits on $B^0$ and $B_s$
annihilation to two photons.

Upper limits on $B^-$ annihilation to a lepton and a neutrino
have been recently published by CLEO-II.~\cite{CLEOln}\
The most stringent upper
limit on the decay to $\tau^-\bar{\nu}_\tau$, which has the smallest
helicity suppression, is set by ALEPH.~\cite{ALEPHln}
CLEO-II also presents
limits on $B^-\to\gamma \lem\bar{\nu}_l$, where
the emission of a photon removes helicity suppression.~\cite{CLEOlng}

All the limits are several orders of magnitudes looser than the values
predicted by the Standard Model. Thus these decays are not likely to
provide useful constraints on the $B$ decay constant nor on the CKM elements
($\Vtd$, $\Vts$, and $\Vub$ are involved) in the near future.
However, these results verify that there are no non-standard couplings
in these processes at the present level of sensitivity.

\section{Inclusive $b\to c$ decays.}

\subsection{Inclusive semileptonic decays,
           $b\to c \lem\bar{\nu}_l$ ($l=e$ or $\mu$).}
\label{sec:isbc}

The inclusive semileptonic branching ratio is related
to a value of $|\Vcb|$.
It also offers tests of QCD models of inclusive $b$ decays, both leptonic
(via its numerator) and non-leptonic (via the denominator).
Finally, it is an important engineering number for the extraction of various
$Z^0$ parameters.

All $b\to cW^-$ decays, hadronic ($W^-\to q_{dn}\bar{q}_{up}$) and
semileptonic ($W^-\to \lem\bar{\nu}_l$), are sensitive to a value $|\Vcb|$.
The main obstacle in the extraction of $|\Vcb|$ from the measured
branching ratios is the difficulty of estimating
strong interaction effects on the decay rate.
Semileptonic decays offer a great simplification compared to hadronic decays
since no strong interactions are involved in the
decay of the virtual $W^-$ and
there is no mixing of lepton generations.
In exclusive semileptonic decay modes strong interactions still play
a crucial role, since a probability for
the creation of a specific hadronic state
containing the $c$ and spectator quarks must be estimated.
For decays summed over all possible hadronic states,
the effect of strong interactions drops out in the lowest order,
and the inclusive rate directly reflects short distance weak interactions.
There are perturbative QCD corrections of the order of
$\alpha_s$ to this approximation.
The lowest non-perturbative correction appears only in the second order
of the expansion in heavy $b$ quark mass.
Thus, the inclusive semileptonic branching ratio
is particularly suitable for
determination of $|\Vcb|$.

Theoretically, $\Gamma(b\to c \lem\bar{\nu}_l)=\gamma_c|\Vcb|^2$,
where in the lowest order $\gamma_c$ can be found from
the muon decay width formula replacing the
lepton masses with the quark masses.
To eliminate the total width from the denominator of the experimental
value of the
branching fraction, a measured $b$ quark lifetime must be used:
$|\Vcb|=\sqrt{\BR(b\to c \lem\bar{\nu}_l)/(\gamma_c\tau_b)}$.
Both experimental quantities in this formula can be measured
very precisely due to their inclusive character.
The lifetime has been measured by experiments at $Z^0$ and in $p\bar{p}$
collisions as discussed in the other paper at this conference.~\cite{Kroll}
The inclusive semileptonic branching ratio has been measured by the
experiments at $\Upsilon(4S)$ and at $Z^0$.
\begin{figure}[htbp]
\vbox{\quad
\vskip-2.2cm
\hbox{\hskip-1cm(a)\hskip-1cm
\lower7.8cm\psfig{file=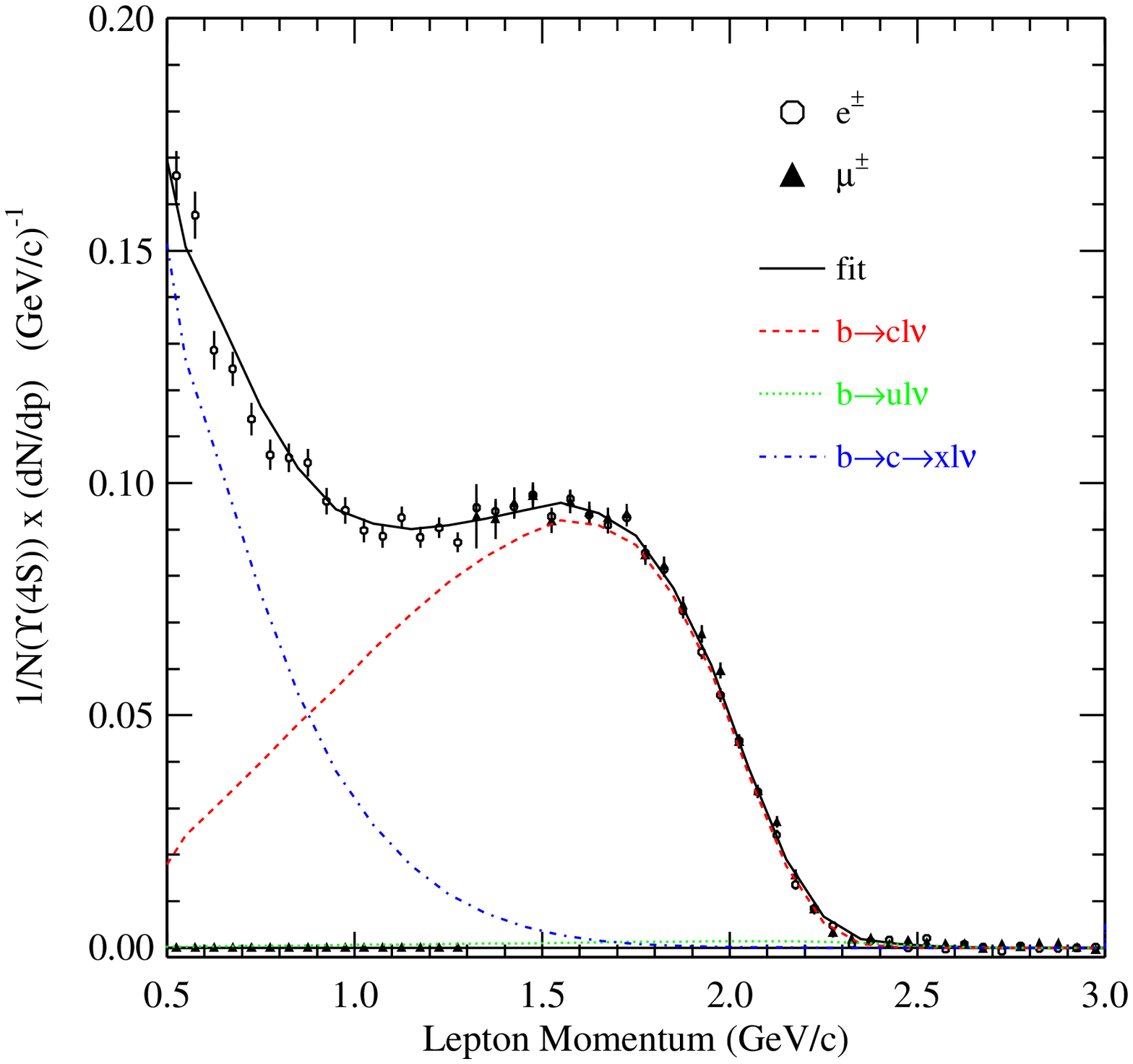,width=10cm,height=10cm}\hskip-0.8cm
(b)\hskip-2.2cm
\lower22.6cm\psfig{file=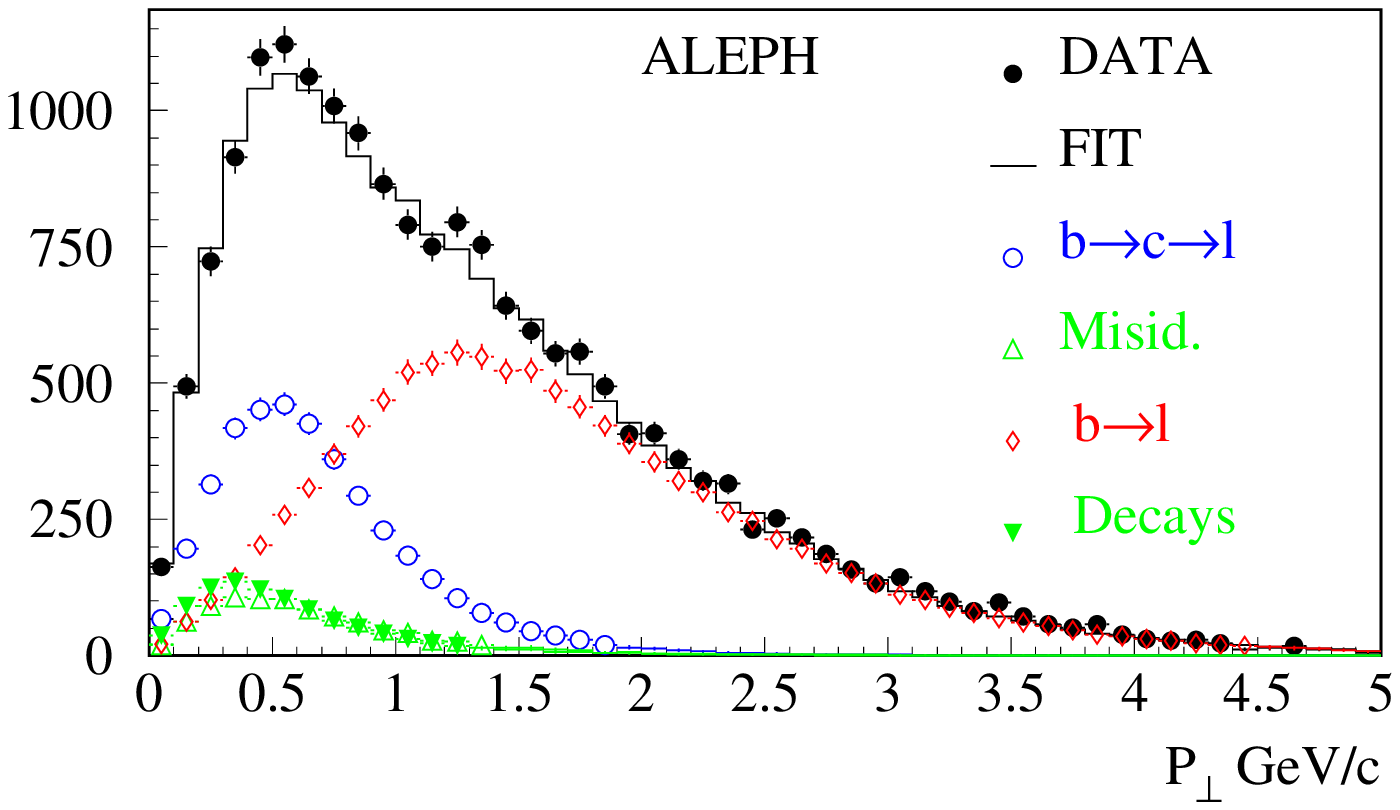,width=11.7cm,height=23.2cm}}
\vskip-16cm
\hbox{\hskip-1cm(c)\hskip-0.4cm
\lower6.6cm\psfig{figure=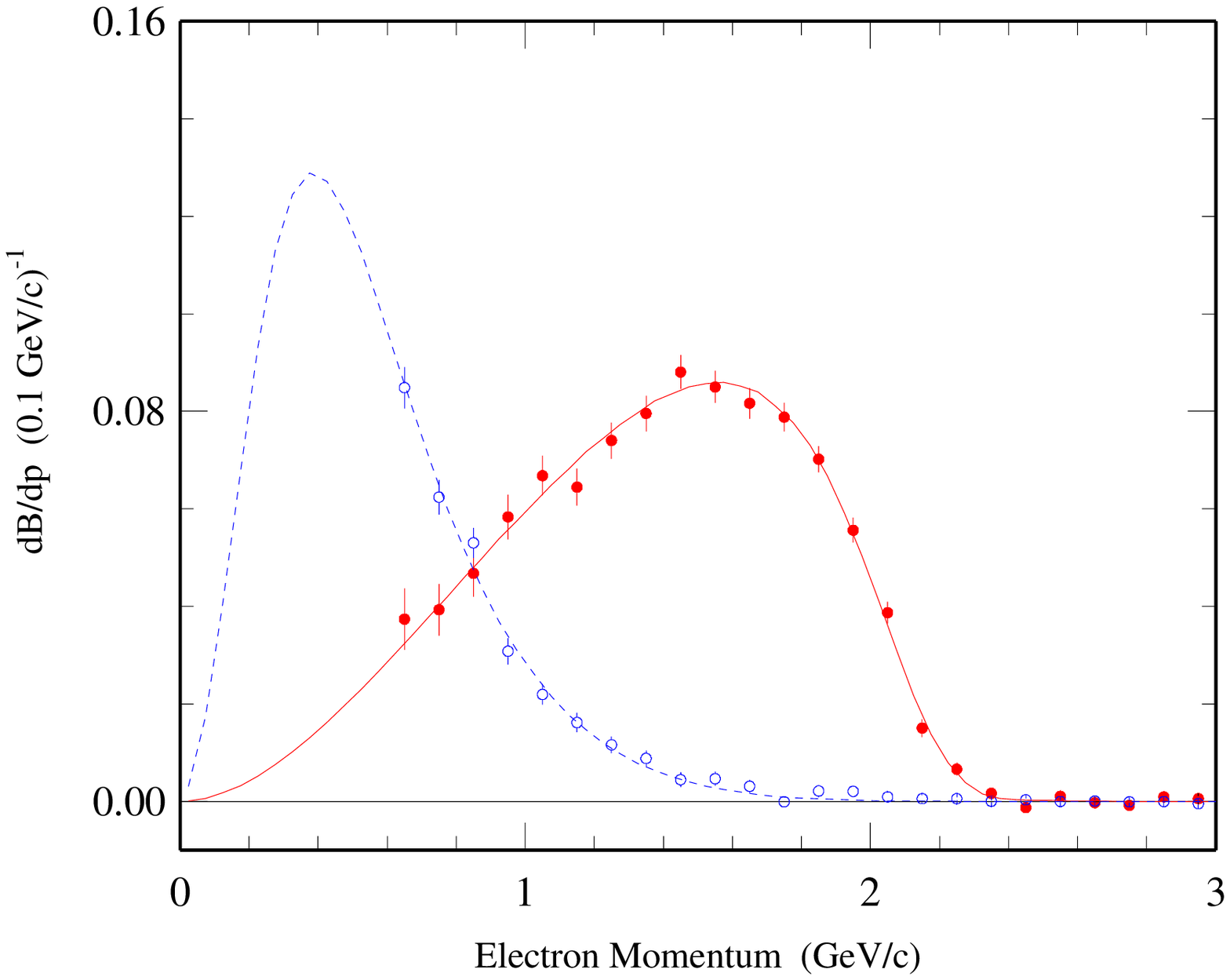,width=8.0cm}
\hskip0.4cm (d) \hskip-2.2cm
\lower11.4cm\psfig{figure=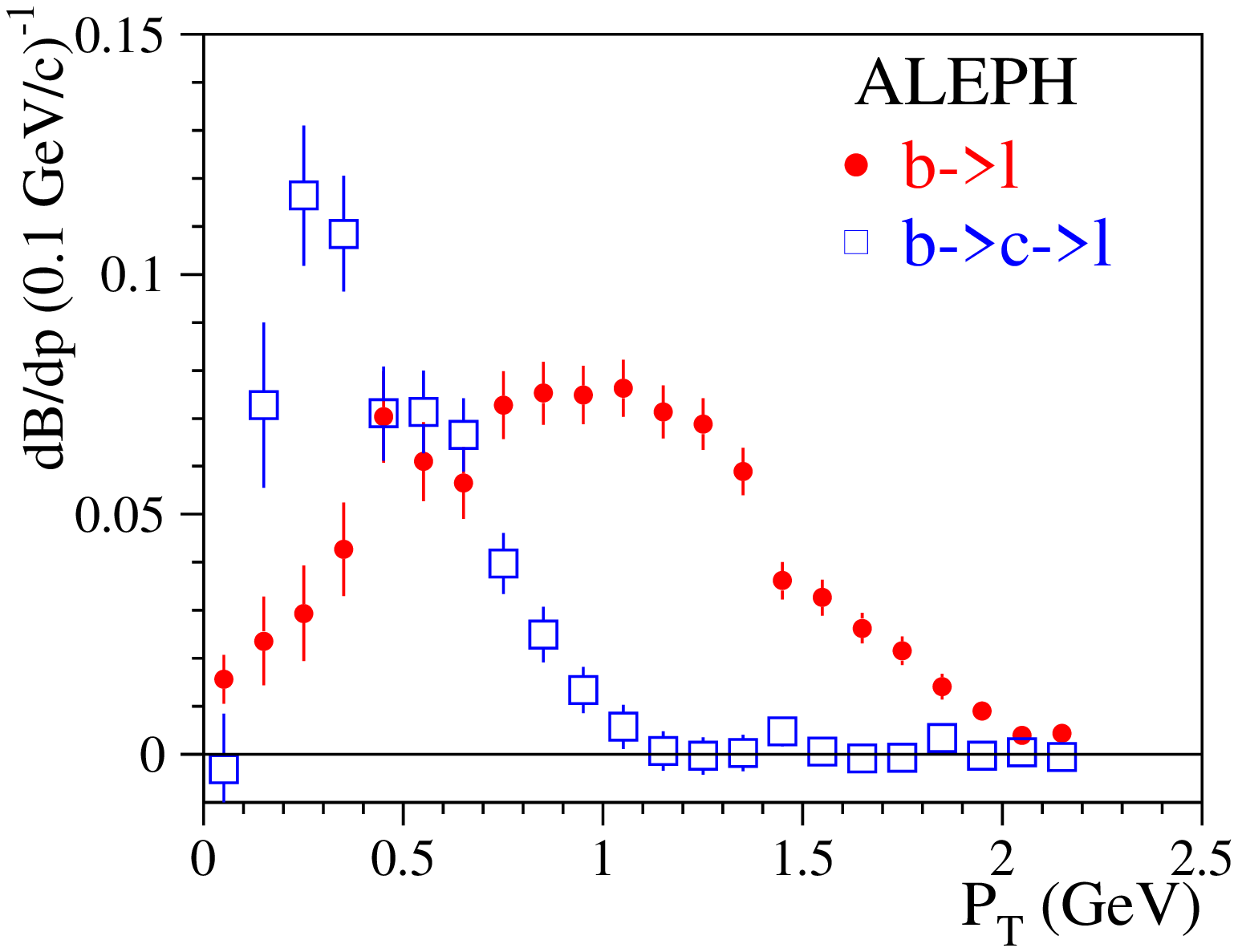,width=11.6cm}}
\vskip-4.7cm
\quad}
\fcaption{Inclusive leptons in $b$ decays.
(a) Lepton momentum spectrum as measured by CLEO-II at $\Upsilon(4S)$.
(b) Spectrum of lepton momentum transverse to the jet direction
    as measured by ALEPH at $Z^0$.
(c) Momentum spectra for prompt leptons, $b\to l$, and
    cascade leptons, $b\to c\to l$, as measured separately in
    dilepton events by CLEO-II.
(d) Transverse momentum spectra for prompt leptons, $b\to l$, and
    cascade leptons, $b\to c\to l$, as measured separately in
    dilepton events by ALEPH.}
\vskip-0.2cm
\label{fig:cleo-aleph}
\end{figure}

Experimentally $\BR(b\to c \lem\bar{\nu}_l)$
is determined as a ratio of $b$ decays producing
a prompt lepton\footnote{Correction for prompt leptons from
the $b\to u \lem\bar{\nu}_l$ decays is small.},
$N_{b\to l}$, and to all $b$ decays, $N_{b\to x}$.
We will first discuss determination of the denominator.
Since non-$B\bar{B}$ decays of $\Upsilon(4S)$ resonance are
expected to be very small,
threshold experiments calculate
$N_{b\to x}$ by counting multi-track events
at and below the $\Upsilon(4S)$ resonance.
At the $Z^0$ peak $N_{b\to x}$ is equal to a number of hadronic
$Z^0$ decays times $R_{b\bar{b}}\equiv \Gamma_{b\bar{b}}/\Gamma_{had}$.
The latter is an interesting quantity by its own as discussed in the
other papers at this conference.~\cite{Renton}
In a simple approach its value can be set to the world average or
to the Standard Model prediction
like in the recent L3 analysis.~\cite{L3isbc}
Using $b$ quark tagging $N_{b\to x}$ can be estimated from
a ratio of number of single-tag events squared and of double-tag events.
When a lepton itself is used to tag $b\bar{b}$ events,
$\BR(b\to c \lem\bar{\nu}_l)$ and $R_{b\bar{b}}$ can be both determined
from single and dilepton samples, however, their values are strongly
anti-correlated. This is the approach taken by most of the previous
measurements at LEP.~\cite{OPALisbc,DELPHIisbc}
In a recent analysis ALEPH~\cite{ALEPHisbc}\
uses a secondary vertex to tag $b\bar{b}$
events and presents practically uncorrelated measurement of these
quantities.

To determine the nominator, $N_{b\to l}$, leptons from other sources
must be estimated.
While  backgrounds from decays of the other
particles, e.g. $N_{c\to l}$, can be suppressed with
$b-$tagging, background from cascade decays of the $b$ quark, $N_{b\to c\to
l}$,
are more difficult to eliminate.
In the $B$ rest frame the prompt leptons have a harder momentum spectrum
than the cascade leptons which allows the discrimination between these
two contributions. Experiments at the $\Upsilon(4S)$ operate almost
at the $B$ rest frame, thus they directly look at the lepton momentum
spectrum.
In $Z^0$ decays, $b$ quarks are produced with a varied momentum
that on average is fairly large ($\sim0.7\frac{m_{Z^0}}{2}$).
This washes out the differences in lepton momenta for the prompt and
the cascade leptons.
Therefore, experiments at $Z^0$ look in the direction in which the
$b$ quarks are not moving, i.e. perpendicularly to their motion
which is approximated by the jet direction.
The distribution of leptons in $P_\perp$ at $Z^0$ is very similar to
the distribution of leptons in $P$ at $\Upsilon(4S)$, as illustrated
in Fig.~\ref{fig:cleo-aleph}a-b.
For $P_{(\perp)}>1.5$ GeV/c the cascade background is very small.
To measure total $b\to l$ rate, the high momentum yield must be
extrapolated to the region dominated by the background.
Alternatively the entire momentum range can be fit to unfold the prompt
and cascade components. In either case, theoretical models
for the $b\to l$ spectrum must be used.
For the most accurate measurements by CLEO-II~\cite{CLEOl}\
at $\Upsilon(4S)$ and
ALEPH~\cite{ALEPHisbc}\ at $Z^0$,
the model dependence is the largest
source of uncertainty.
These two measurements dominate world
average values of $\BR(b\to c \lem\bar{\nu}_l)$ at
$\Upsilon(4S)$~\cite{bl4swa}, $(10.56\pm0.17\pm0.33)\%$,
and at $Z^0$~\cite{blz0wa}, $(10.89\pm0.18\pm0.24)\%$.
The first error is experimental (statistical and systematic)
and the second error is an estimate of the model dependence.
The central values of the branching fractions correspond
to the choice of the quark level model by Altarelli et al.~\cite{ACCMM}\
to predict the shape of the $b\to c \lem\bar{\nu}_l$ lepton spectrum.
The model dependence was estimated as a difference
between the meson level model by Isgur et al.~\cite{ISGW}
and the quark level model.

Model independent measurements can be obtained using a technique
developed by the ARGUS experiment that utilizes charge and angular
correlations in dilepton events.~\cite{ARGUSdil}
Selection of a high momentum lepton tags $b\bar{b}$ events.
At $\Upsilon(4S)$, the $b$ quarks are produced almost at rest, thus
a lepton from the cascade decay
of the same $b$ quark
should be approximately acollinear to the tag lepton and
can be eliminated by the angular cut.
The prompt and
the cascade lepton from the other $b$ quark decay
can be distinguished by electric charge.
Neglecting $B\bar{B}$ oscillation, the prompt lepton
should have an opposite charge to the tag lepton.
Since the prompt and the cascade spectra are measured
independently (except for small feed-across),
the prompt component is measured almost in full momentum range
and the model dependence is eliminated.
The CLEO-II experiment applied the ARGUS technique to a larger
data sample (Fig. \ref{fig:cleo-aleph}c).~\cite{CLEOdil}\
An average of these two experiments is
$(10.16\pm0.38\pm0.11)\%$ in good agreement with the model
dependent determinations.
ALEPH presents an analysis that adopts this technique to the LEP
data.~\cite{ALEPHisbc}
Since $b$ quarks produced at $Z^0$ are fast,
the angular correlation is reversed, i.e.
the high  $P_{\perp}$ lepton,  used as a tag, and
the cascade lepton from
the decay of the same $b$ quark
are collinear (Fig. \ref{fig:cleo-aleph}d).
As in the case of the CLEO measurement, the model dependence is diminished
at the expense of the other errors: $(11.01\pm0.36\pm0.11)\%$.

The measurements at $Z^0$ are somewhat larger than at $\Upsilon(4S)$.
Since the average $b$ quark lifetime at $Z^0$ is smaller than
at $\Upsilon(4S)$~\cite{Kroll}
(no baryons are produced at $\Upsilon(4S)$)
the opposite effect was expected.
Decay widths, $\Gamma=\BR/\tau$,
measured at $\Upsilon(4S)$\footnote{For the best estimate
of the semileptonic branching fractions, we averaged
the results from single lepton and dilepton samples,
$\BR(B\to X_c \lem\bar{\nu}_l)_{\Upsilon(4S)}=(10.37\pm0.20\pm0.23)\%$.
The second error expresses the model dependence.
To correct for the lifetime,
we used $(\tau_{B^-}+\tau_{B^0})/2=1.60\pm0.03$ ps.~\cite{Kroll}},
$64.8\pm1.7\pm1.4$ ns$^{-1}$ (the second error expresses the model
dependence),
and at $Z^0$\footnote{We did not include the model independent
measurement by ALEPH because of a strong correlation to their other
result that has a smaller overall error.~\cite{ALEPHisbc}
To correct for the lifetime, we used $\tau_{b}=1.55\pm0.02$ ps.~\cite{Kroll}},
$70.2\pm1.5\pm1.5$ ns$^{-1}$,
differ by about two standard deviations.
Fortunately, this disagreement is small compared to theoretical
uncertainties in the $\gamma_c$
factor:~\cite{isbcgamma,Neubert} $42.4\pm8$ ps$^{-1}$.
Therefore, for $|\Vcb|$ determination we average the
$\Upsilon(4S)$ and $Z^0$ measurements and
cover the disagreement by increasing the experimental error:
$\Gamma(b\to c \lem\bar{\nu}_l)=67.3\pm2.7$ ns$^{-1}$.
That leads to:
$$ |\Vcb|=0.0398\pm0.0008(experimental)\pm0.0040(theoretical).$$
Theoretical uncertainties due to higher order
perturbative corrections are somewhat controversial at present.
Following Neubert~\cite{Neubert}\ we used above the conservative
estimate. Bigi et al.\ argue that the theoretical error is a factor
of two smaller.~\cite{BigiPisa}

\subsection{Charm counting in bottom decays.}

Value of the semileptonic branching ratio measured for
$B$ mesons, $(10.4\pm0.3)\%$,
is low compared to the theoretical predictions that
typically give $\ge12.5\%$.~\cite{bigi}
It has been recently suggested that lower predictions can be obtained
by enhancement of the $b\to c(\bar{c}s)$ contribution to the total
width.~\cite{ccs}
At the same time the predicted number of charm
quarks per $b$ quark decay ($n_c$) increases.
To accommodate the experimental value of the semileptonic
branching ratio $n_c$ must be raised from
about $n_c=1.15$ to about $1.30$\%~\cite{hitoshi}\
in contradiction with the most recent measurements of inclusive
branching ratios by CLEO-II,
$n_c<(1.16\pm0.05)$.
The contributing measurements are shown in Table~\ref{tab:CLEOcc}.
Thus, discrepancy between the theory and the experiment remains to be
resolved.
\begin{table}[hbtp]
\tcaption{Inclusive charm branching ratios in $B$ meson decays
as measured by CLEO-II. The first two numbers are preliminary.}
\label{tab:CLEOcc}
\small
\def\1{\phantom{<2\times(}}
\def\2{\phantom{<2\times}}
\def\3{\phantom{<}}
\begin{center}
\begin{tabular}{||c|l|r||}
\hline\hline
{} & {} & {} \\
$B\to X+\dots$ & \multicolumn{1}{c|}{$n_c$} & \multicolumn{1}{c||}{Ref.} \\
{} & {} & {} \\
\hline
{} & {} & {} \\
$D^0$ &         $\1 0.645\pm0.031$ & ~\lowcite{Moneti}~ \\
$D^+$ &         $\1 0.235\pm0.029$ & ~\lowcite{Moneti}~ \\
$D_s$ &         $\1 0.121\pm0.017$ & ~\lowcite{CLEODsX}~ \\
$\Lambda_c$ &   $\1 0.064\pm0.011$ & ~\lowcite{CLEOLc}~ \\
$\Xi^0_c$ &     $\1 0.024\pm0.013$ & ~\lowcite{CLEOSc}~ \\
$\Xi^+_c$ &     $\1 0.015\pm0.007$ & ~\lowcite{CLEOSc}~ \\
$\psi$ &       $\3 2\times(0.0080\pm0.0008)$ &  ~\lowcite{CLEOcc}~ \\
$\psi'$ &      $\3 2\times(0.0034\pm0.0005)$ &  ~\lowcite{CLEOcc}~ \\
$\chi_{c1}$ &  $\3 2\times(0.0037\pm0.0007)$ &  ~\lowcite{CLEOcc}~ \\
$\chi_{c2}$ &  $\3 2\times(0.0025\pm0.0010)$ &  ~\lowcite{CLEOcc}~ \\
$\eta_c$ &     $\!<\!2\times\phantom{(}0.009$ & ~\lowcite{CLEOcc}~
 \\
{} & {} & {} \\
\hline
{} & {} & {} \\
Total &    $\!\!\2 <\!1.16\pm0.05$  & {} \\
{} & {} & {} \\
\hline\hline
\end{tabular}
\\[-1cm]\quad
\end{center}
\end{table}

\subsection{Inclusive semileptonic decays,
           $b\to c \tau^-\bar{\nu}_\tau$.}

\vbox{
Rate for semileptonic decays to the $\tau$ lepton is expected to be
smaller than to electron or muon because of the phase space suppression.
Since the $\tau$ lepton decays to many different channels and
produces at least one additional neutrino, this rate is difficult
to measure experimentally.
The yield of events with large missing energy in a hemisphere
defined by the jet direction opposite to the vertex or lepton $b$ tag
in $Z^0$ decays is a sensitive
measure of the semileptonic rate for decays to $\tau$'s.
There are new measurements of this rate by L3~\cite{L3tauX}\
and OPAL~\cite{OPALtauX}, in addition
to previous determinations by L3~\cite{L3tauXold}
and ALEPH.~\cite{ALEPHln}
The world average value,
 $\BR(b\to c\tau^-\bar{\nu}_\tau)=(2.6\pm0.3)\%$,
is in good agreement with the Standard Model predictions,
$2.3-2.8$\%.~\cite{thtauX}
This agreement can be used to set a lower limit on charged Higgs mass
in Two-Higgs-Doublet-Models, $m_{H^-}>2\tan\beta$ GeV,~\cite{ALEPHln}
that dominates over the limit obtained from $b\to s\gamma$ decays for
very large values of $\beta$ (see Sec.~\ref{sec:bsg}).
\vskip-5cm
}
\newpage

\section{Exclusive semileptonic decays,
           $B\to X_c \lem\bar{\nu}_l$, $X_c=D, D^*, D^{**}, \dots$}

\subsection{Form factors}

Unlike for inclusive semileptonic decays,
weak and strong interactions cannot be easily separated
for exclusive decays.
The decay width still carries
information about $|\Vcb|$,
$\Gamma(B\to X \lem\bar{\nu}_l)=\gamma_X|\Vcb|^2$,
but the theoretically predicted width $\gamma_X$ crucially
depends on the effect of the strong interactions which
is parametrized into form factors ($FF$) that are a function of
$q^2$ i.e. four-momentum transfer between the initial
and final state mesons, $(P_B^\nu-P_X^\nu)^2$;
$\gamma_X=G^2_F\int^{(m_B-m_X)^2}_0 \dots FF(q^2) \dots dq$.
The form factors contain information about structure of initial
and final state mesons.
For decays to pseudo-scalars and negligible lepton
mass there is only one form factor, $F_1(q^2)$.
Three form factors are involved in decays to vector mesons,
$A_1(q^2)$, $A_2(q^2)$, $V(q^2)$.
Phenomenological models are used to predict form factors.
The models must predict both $q^2$ dependence and absolute
normalization of the form factors.
While the $q^2$ dependence can be eventually verified on the
data (large statistics are needed), the normalization
must be taken from the models.
Therefore, measurements of exclusive semileptonic branching
fractions lead to model dependent determinations of $|\Vcb|$.

\subsection{Heavy Quark Effective Theory}

\def\fv{v}
\def\iw{{\cal F}}
In recent years, development of the Heavy Quark Effective Theory
allowed overcoming most of the model dependence in extraction of $|\Vcb|$
from the exclusive semileptonic decays.~\cite{hqet}\
In the limit of infinitely heavy quark mass, heavy-light mesons
behave like single electron atoms.
Spin of the heavy quark (\lq\lq nucleus'')
can be to a good approximation neglected.
Also, different flavor mesons (\lq\lq isotopes'')
have the same structure when
expressed in heavy quark four-velocity, $\fv$.
In this approximations, $B$, $D$ and $D^*$ mesons have the same
structure and there is only one universal form factor
called the Isgur-Wise function~\cite{iw}, $\iw$,
that depends on $q^2$ via four-velocity transfer, $(\fv_B-\fv_X)^2$;
$$
\begin{array}{c}
F_1(q^2) = \frac{m_B+m_X}{2\,\sqrt{m_B\,m_X}} \iw(y(q^2))\\
 V(q^2) =A_2(q^2) =\frac{ A_1(q^2) }{1-q^2/(m_B+m_X)^2} =
\frac{m_B+m_X}{2\,\sqrt{m_B\,m_X}} \iw(y(q^2))\\
y(q^2)=\frac{m_B^2+m_X^2-q^2}{2\,m_B\,m_X}=\fv_B\cdot\fv_X\equiv
\fv_b\cdot\fv_c = 1 - (\fv_B-\fv_X)^2/2
\end{array}
$$
The heavy quark symmetry limit is an approximation to QCD.
To make HQET useful in practice, $\alpha_s$,
$\Lambda_{QCD}/m_b$, and $\Lambda_{QCD}/m_c$ corrections must be
calculated.

General form of the universal form factor, $\iw(y)$, is not predicted
by HQET and phenomenological models are still needed for calculation
of the total expected rate.
However, at one point at which there is no recoil between the
initial and final state mesons, $\fv_X=\fv_B$ (i.e. at $y=1$ or
at maximal $q^2$),
the hadronic system is unaffected, and an overlap
between $B$ and $X$ wave functions (that are identical) is complete:
$\iw(1)=1$. This is a model independent prediction.
Measurement of the exclusive decay rate at this particular point
allows an almost model independent determination of $|\Vcb|$.

\subsection{Exclusive $B\to D \lem\bar{\nu}_l$.}
\label{sec:d0lnu}

So far the
decays $B\to D \lem\bar{\nu}_l$ have been studied only at $\Upsilon(4S)$.
The main background originates from the other semileptonic
decays, $B\to D^* \lem\bar{\nu}_l$ and $B\to D^{**} \lem\bar{\nu}_l$.
Soft pion from the $D^*\to\pi D$ decay can easily escape detection.
Since the neutrino is also undetected, the main experimental challenge
is to distinguish events with a missing neutrino from events in
which more particles have been missed.
ARGUS~\cite{ARGUSdlnu}\
and CLEO-I.5~\cite{CLEOdlnuold}\
used a technique in which the neutrino mass was
calculated in an approximate way:
$m_\nu^2 
         = (E_B-E_{Dl})^2-|\vec{p}_B|^2 - |\vec{p}_{Dl}|^2  +
  2\,|\vec{p}_B|\,|\vec{p}_{Dl}|\,\cos\theta_{B,Dl}
  \approx(E_{beam}-E_{Dl})^2-(E_{beam}^2-m_B^2) - |\vec{p}_{Dl}|^2$.
Neglecting the term with unknown $\theta_{B,Dl}$ angle
does not change the mean value of $m_\nu^2$ but adds to the experimental
resolution.
Within this resolution the $B\to D \lem\bar{\nu}_l$ signal
peaking at zero overlapped with the background peaks only slightly
shifted upwards in the $m_\nu^2$ variable.
The fit of the signal and background contributions to the  $m_\nu^2$
distribution determined the $B\to D \lem\bar{\nu}_l$ branching ratio with
rather large errors, especially for  $B^-\to D^0 \lem\bar{\nu}_l$
where the $D^*$ background peak was much larger than the signal peak.

Recent CLEO-II analysis~\cite{CLEOdlnunew}\
overcomes this background limitation at the expense of
statistics by
reconstructing neutrino four-momentum:
$(E_\nu,\,\vec{p}_\nu)\approx(2\,E_{beam}-E_{visible}^{~event},
-\vec{p}_{visible}^{~event})$.
Charge conservation is imposed to eliminate events with
unreconstructed charged tracks.
To suppress events with more than one neutrino
only one lepton in the event is allowed.
Cut on neutrino mass, $E_\nu^2-p_\nu^2$, reduces the background with
other unreconstructed particles e.g. $K^0_L$.
Since the achieved resolution is better for the missing momentum
($\sigma(p_\nu)=0.11$ GeV/c) than for the missing energy
($\sigma(E_\nu)=0.26$ GeV) the former is used to impose
the beam energy constraint ($E_{Dl}+|\vec{p}_\nu|=E_{beam}$) and
to calculate the $B$ meson mass ($M_B=\sqrt{E_{beam}^2-
(\vec{p}_{Dl} + \vec{p}_\nu)^2}$).
The $D^*$ background is significantly reduced with this method, thus
$\BR(B^-\to D^0 \lem\bar{\nu}_l)$ is measured with improved precision.

Averaging the previous ARGUS measurement,
$\BR(\bar{B}^0\to D^+ \lem\bar{\nu})=(2.0\pm0.7\pm0.5)\%$,
and the new CLEO-II measurement,
$\BR(B^-\to D^0 \lem\bar{\nu})=(2.0\pm0.3\pm0.5)\%$,
corrected for the $B$ lifetimes\footnote{
We used $\tau_{B^-}=1.62\pm0.05$ ps
and $\tau_{B^0}=1.57\pm0.05$ ps.~\cite{Kroll}}\
we obtain:
$\Gamma(B\to D \lem\bar{\nu}_l)=12.4\pm3.0$ ns$^{-1}$.
{}From several models of form factors~\cite{FFdlnu}\
we estimate, $\gamma_D=(8.9\pm3.1)$ ps$^{-1}$, where the error
is the difference between the largest and the smallest prediction.
The above leads to,
$$ |\Vcb|=0.0373\pm0.0045\pm0.0065 $$
where the first error is experimental and the second reflects
the model dependence.

\begin{figure}[tbhp]
\vbox{
\quad\vskip-6.5cm
\hbox{
\psfig{file=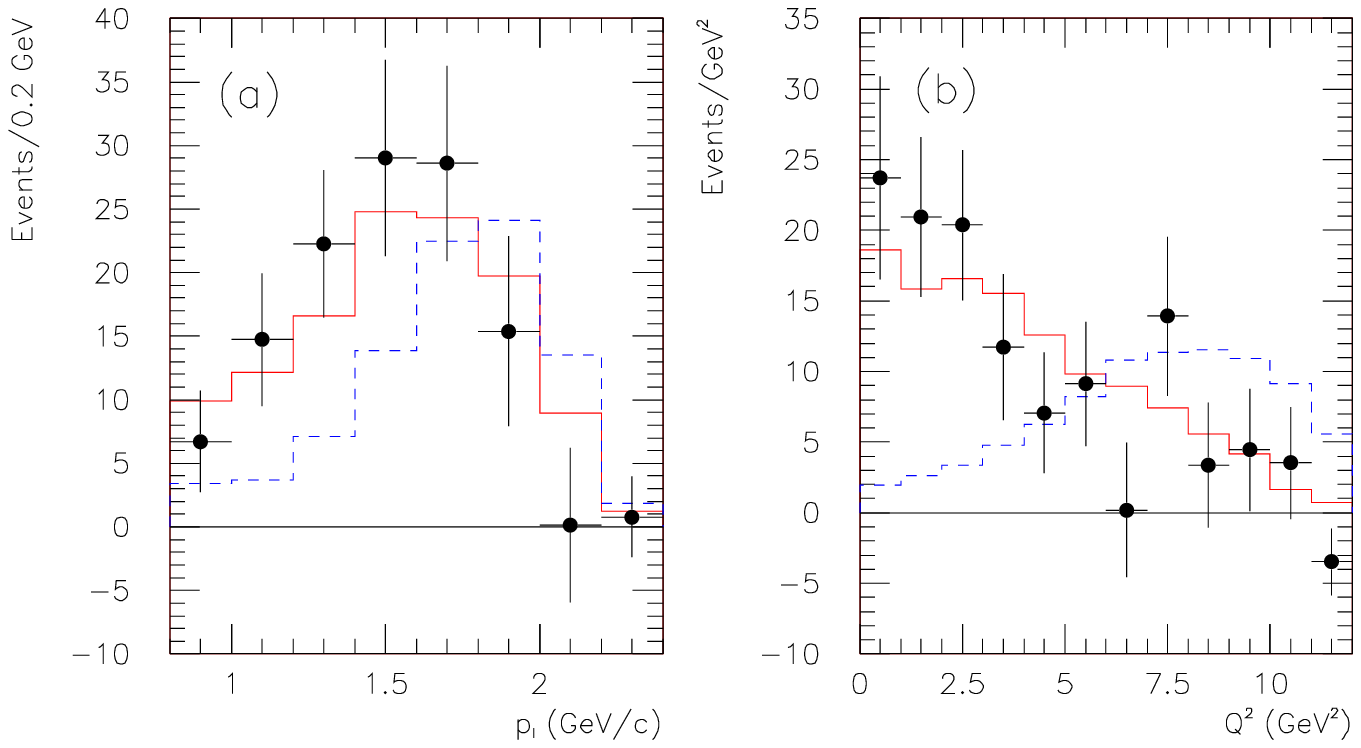,width=20cm,height=15cm}
}
\vskip-5.0cm\quad}
\fcaption{Lepton momentum (a) and $q^2$ (b) distributions in
$B^-\to D^0 \lem\bar{\nu}_l$ decays measured by CLEO-II (points with
error bars). The data points are not efficiency corrected.
The solid curves correspond to the Monte Carlo simulation of
the signal, whereas the dashed curves represent the Monte Carlo
simulation of the $B\to D^* \lem\bar{\nu}_l$ background.}
\label{fig:CLEOd0lnu}
\vskip-0.5cm\quad
\end{figure}

Fig.~\ref{fig:CLEOd0lnu}\ shows $q^2$ distribution
measured for $B^-\to D^0 \lem\bar{\nu}$ by CLEO-II.
As expected from the helicity suppression in $B$ decay to a pseudo-scalar,
the rate quickly tends to zero when approaching the maximal $q^2$
value (zero recoil point). At the same time the $D^*$ background
increases.  This illustrates that the  $B^-\to D^0 \lem\bar{\nu}$
channel will be very difficult to use to determine $|\Vcb|$
in model independent way even for much larger experimental statistics.
This channel is also disfavored theoretically as the $\Lambda_{QCD}/m_c$
correction to the heavy quark symmetry limit does not vanish.

\subsection{Exclusive $B\to D^* \lem\bar{\nu}_l$.}

\subsubsection{Branching ratio and model dependent determination
               of $|\Vcb|$.}

The technique that uses $m_\nu^2$ determined in
the approximate way to extract an exclusive semileptonic
branching fraction works better for the $D^*$ channel than for
the $D$ channel, since a ratio of the background ($D^{**}$)
to the signal branching ratios is favorable.
In a recently published analysis,~\cite{CLEOIIdsvcb}
CLEO-II modified
this technique and used  a two-dimensional distribution of
$m_\nu^2$ vs. $2\,|\vec{p}_B|\,|\vec{p}_{D^2l}|$
(the latter is a factor multiplying the unknown $\cos\theta_{B,D^*l}$
term - see Sec.~\ref{sec:d0lnu}).

Since a slow pion from the $D^*\to\pi D$ decay preserves the $D^*$
direction in the laboratory system and its momentum is proportional
to the $D^*$ momentum, the entire $D^*$ four-momentum can be approximately
reconstructed by detecting the transition pion alone.
This method allowed ARGUS~\cite{ARGUSdspart}
to overcome the statistical limitation of
their original measurement.~\cite{ARGUSds}

This year there are two new measurements of
the $\bar{B}^0\to D^{*0} \lem\bar{\nu}_l$ branching
ratio presented by the LEP experiments: ALEPH~\cite{ALEPHdsvcb}\
and DELPHI.~\cite{DELPHIdsvcb}\
At $Z^0$ the slow pion is boosted up in momentum into a region
of large and uniform detection efficiency.
This advantage over the measurements at the $\Upsilon(4S)$ is
offset by an additional normalization error from the $Z^0\to B^0$ rate
and by reconstruction difficulties due to unknown $B^0$ energy and
momentum. The latter has been overcome with help of silicon vertex
detectors that are able to determine direction of the $B^0$
momentum from the separation of the $B^0$ decay
vertex from the interaction point. Neutrino energy is calculated
from the missing energy in the $D^*l$ hemisphere:
$E_\nu\approx E_{beam}-E_{visible}^{~hem.}$.
These two quantities, together with the four-momentum conservation and
the mass constraints allow full reconstruction of the
$\bar{B}^0\to D^{*0} \lem\bar{\nu}_l$ kinematics.

For most of the branching ratio measurements uncertainty in the
$D^{**}$ background is the dominant error.
A cut on the neutrino mass is not as effective in
suppression of the $D^{**}$ background at $Z^0$ as it is at $\Upsilon(4S)$.
However, vertexing again comes handy.
Events with more than one charged pion pointing to the $B^0$ decay
vertex are rejected.
The $D^{**}$ background rejection achieved by ALEPH is a factor of
2.5 worse than in the CLEO-II analysis but comparable to the
ARGUS results. DELPHI has a higher background partly because the
partial reconstruction of $D^*$'s is used.

Averaged over all determinations,
$\BR(\bar{B}^0\to D^{*+} \lem\bar{\nu}_l)=(4.47\pm0.25)\%$,~\cite{dspav}\
$\BR(B^-\to D^{*0} \lem\bar{\nu}_l)=(5.13\pm0.67)\%$,~\cite{ds0av}
and $\Gamma(B\to D^* \lem\bar{\nu}_l)=28.8\pm1.7$ ns$^{-1}$.
{}From the model predictions:~\cite{FFdlnu}\
$\gamma_{D^*}=23.6\pm8.2$ ps$^{-1}$.
Thus,
$$ |\Vcb|=0.0350\pm0.0010\pm0.0061 $$
where the second error reflects the model dependence.

\subsubsection{Rate at zero recoil point and model independent determination
               of $|\Vcb|$.}

The  $B\to D^* \lem\bar{\nu}_l$ channel is favored for model independent
determination of  $|\Vcb|$ both theoretically and experimentally.
It has been shown that the
first oder non-perturbative correction to the
heavy quark symmetry limit must be zero.~\cite{Luke}
The second order
non-perturbative correction and the
first order perturbative correction
have been estimated by several authors,
$\iw(1)=0.91\pm0.04$.~\cite{Neubert}
Experimentally, the $D^*$ channel has the largest branching fraction
among all exclusive semileptonic decays.
Nevertheless, statistics near the zero recoil point
is still very limited, thus the experiments
use full $y$ range  to extrapolate the measured rate
to the $y=1$ point.
Within the present statistical errors the data are well
described by a linear function,
$|\Vcb|\iw(y)=|\Vcb|\iw(1)\,(1-\hat\rho^2(y-1))$,
where $|\Vcb|\iw(1)$ and $\hat\rho^2$ are free parameters.
Various formulae were tried by some experiments to evaluate
the extrapolation error.

\begin{figure}[tbhp]
\vbox{
\quad\vskip-2.0cm
\hbox{(a) \hskip-2cm
\lower11.0cm\psfig{file=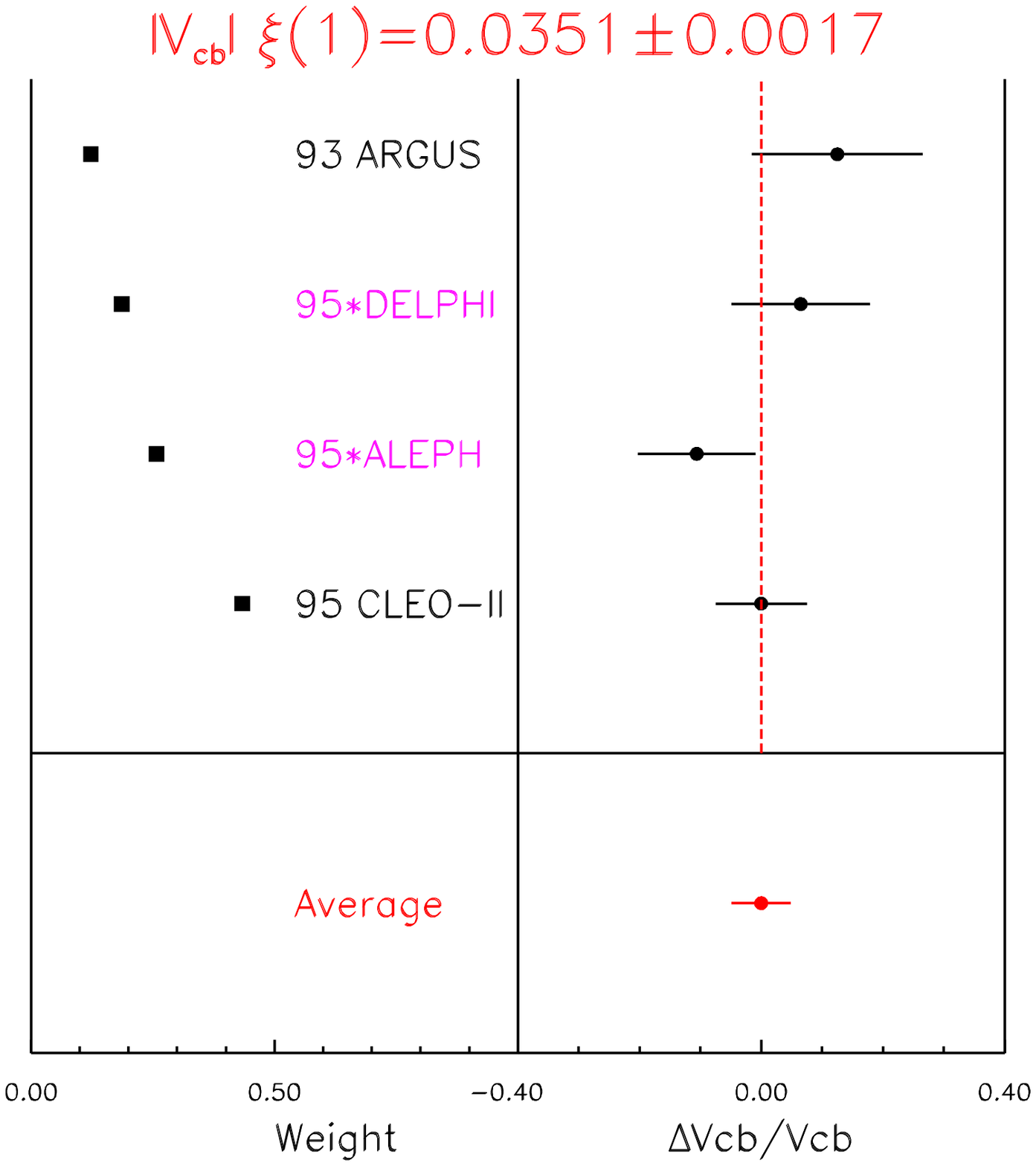,width=10cm}\hskip-1cm
(b) \hskip-2cm
\lower11.0cm\psfig{file=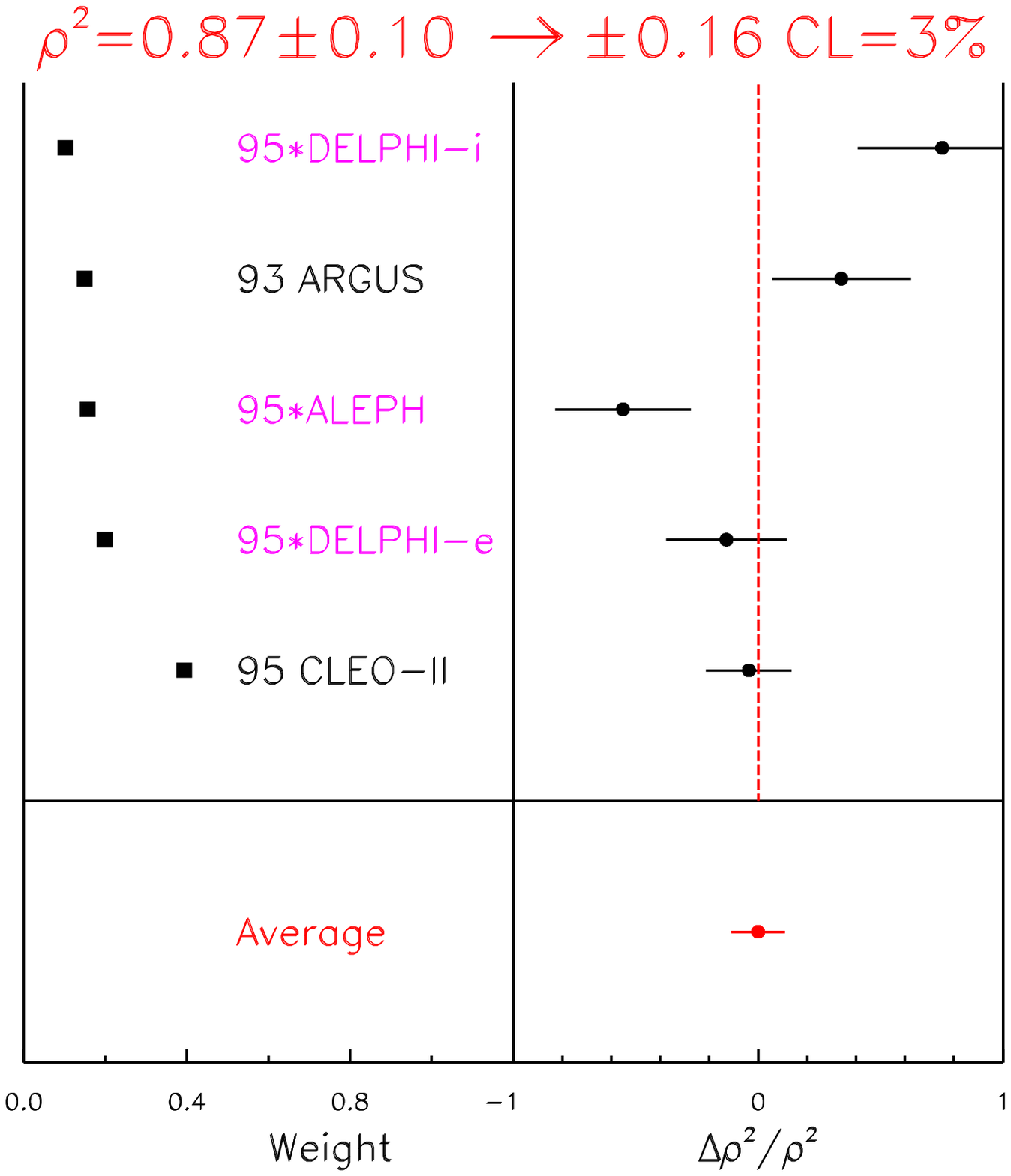,width=10cm}
}
\vskip-2.5cm\quad}
\fcaption{Comparison of various measurements of:
\quad (a) $|\Vcb|\iw(1)$;\quad (b) slope of $\iw(y)$.
Relative deviations of each measurement from the average value are
shown. Weights in which each experiment contributes to the average are
also shown. Preliminary measurements are indicated by a star.
Two measurements of the slope parameter from DELPHI
correspond to the partial (-i)
and full (-e) $D^*$ reconstruction methods.}
\label{fig:vcbf}
\vskip-0.5cm\quad
\end{figure}
The above method was first used by ARGUS.~\cite{ARGUSds}
CLEO-II recently published
results based on much higher statistics and improved
event selection.~\cite{CLEOIIdsvcb}\
Also ALEPH~\cite{ALEPHdsvcb}\
 and DELPHI~\cite{DELPHIdsvcb}\
 have applied this method and report new
results.
Systematic effects in the measurements
at $\Upsilon(4S)$ and at $Z^0$ are quite different.
Thanks to the
rest frame kinematics in the $\Upsilon(4S)$ experiments, the
$y$ variable is determined by neglecting a small term
that depends on $\theta_{B,D^*}$ angle,
$y = v_B\cdot v_{D^*} = \frac{E_B}{m_B} \frac{E_{D^*}}{m_{D^*}} -
\frac{|\vec{p}_B|}{m_B} \frac{|\vec{p}_{D^*}|}{m_{D^*}}
\cos\theta_{B,D^*} \approx \frac{E_{D^*}}{m_{D^*}}$.
The achieved resolution is about 4\%\ of the total range of $y$
variation.
In $Z^0$ decays, both terms are large and $y$ resolution crucially
depends on the measurement of the small $\theta_{B,D^*}$ angle with
help of silicon vertex detectors.
The resolution achieved at LEP is 14-18\%.
Uncertainties in unfolding the resolution effects becomes the
dominant systematics for the DELPHI measurement.

The results are summarized in Fig.~\ref{fig:vcbf}.
While there is a good agreement among all measurement of
$|\Vcb|\iw(1)$, probability that all slope measurements are
consistent is only 3\%.
Ignoring this disturbing fact, an average over all experiments
gives $|\Vcb|\iw(1)=0.0351\pm0.0017$.
This result relies on the assumption that the form factor $\iw(y)$
is linear.
The effect of using the other functional forms was investigated
by ARGUS~\cite{ARGUSds}\ and by
Stone for the CLEO-II data.~\cite{Stone}
The latter study, based on the higher statistics, shows that
the obtained value of $|\Vcb|\iw(1)$ may change up to $+4\%$.
The variation is asymmetric because the assumed functions have
a positive curvature as expected when approaching a pole.
We include this uncertainty in the experimental error:
$$ |\Vcb|=0.0386_{-0.0019}^{+0.0025}(experimental)\pm0.0017(theoretical).$$

\subsubsection{Tests of the Heavy Quark Effective Theory.}
\begin{figure}[hbtp]
\vbox{
\quad\vskip-5cm
\hbox{\hskip-1.2cm\psfig{file=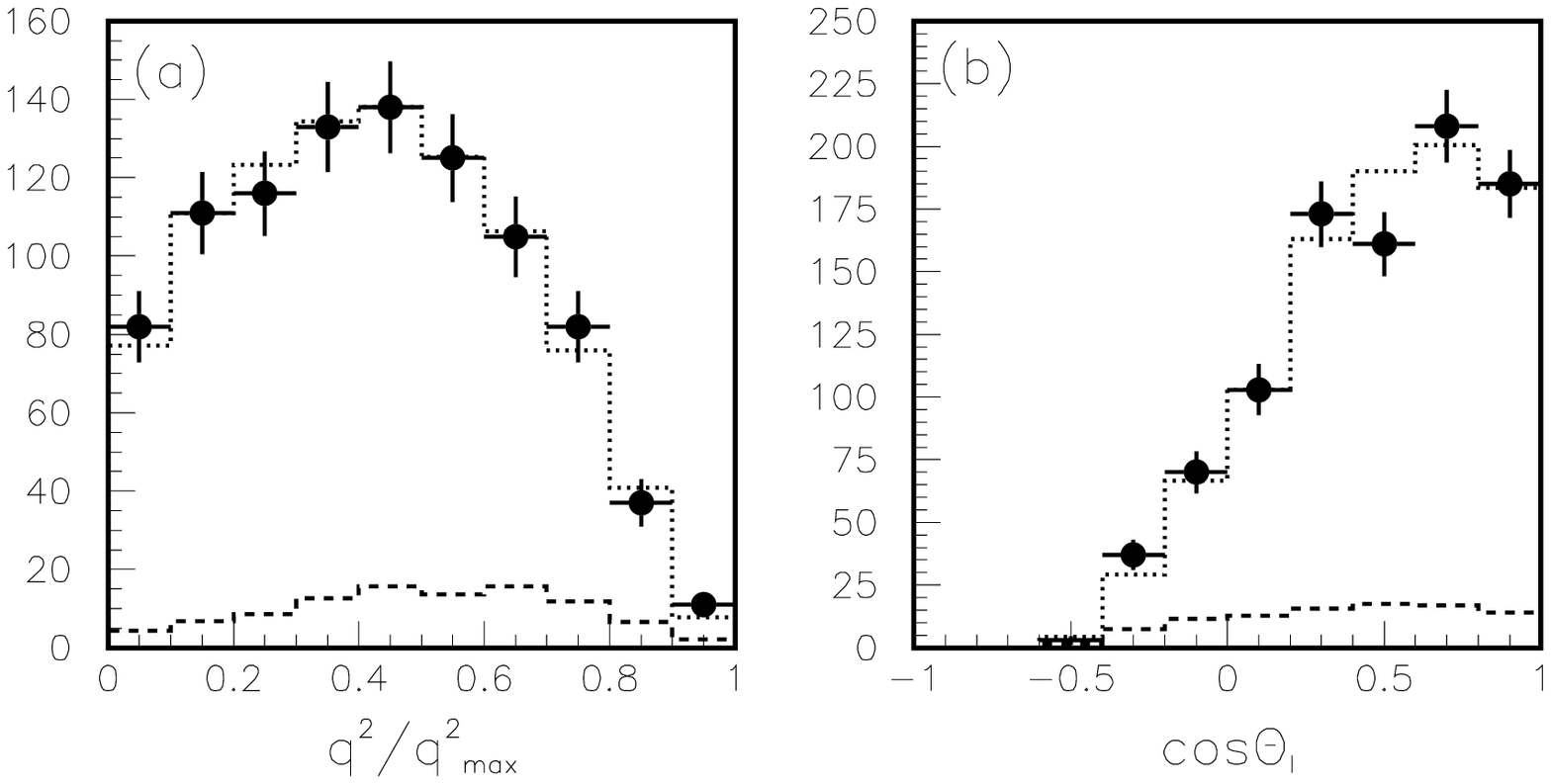,width=20cm,height=15cm}}
\vskip-10.0cm
\noindent
\hbox{\hskip-1.2cm\psfig{file=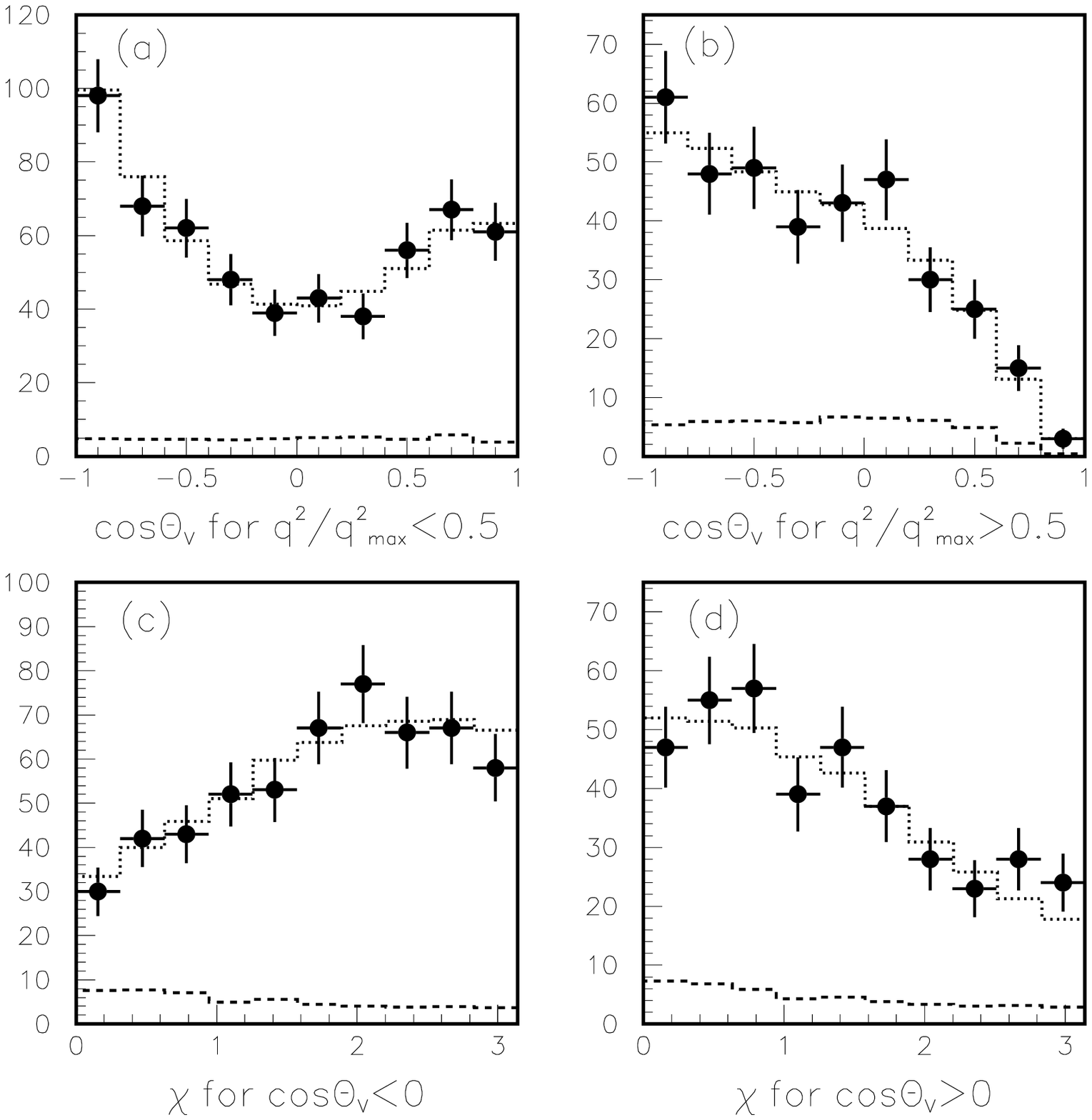,width=20cm,height=15cm}}
\vskip-0.8cm\quad}
\fcaption{Distribution of $q^2$ and various angles
(see Ref.~\lowcite{CLEOhqet}
for definitions) describing $B\to D^* \lem\bar{\nu}_l$
decay as measured by CLEO-II. The points with error bars are the data.
The dotted histograms show the results of the fit of the HQET motivated
form factors. The dashed lines show the estimated backgrounds included
in the fit.}
\label{fig:CLEOhqet}
\vskip-0.5cm\quad
\end{figure}

Since we rely on HQET for $|\Vcb|$ determination from the measured
rate at the zero recoil point, it is important to test the other
prediction of HQET, namely that there should be only one
form factor in the heavy quark symmetry limit.
Since in the real world this symmetry is broken,
it is appropriate to test the modified relation among
the form factors:
$$
\frac{V(q^2)}{R_1(y(q^2))} =
\frac{A_2(q^2)}{R_2(y(q^2))}
=\frac{A_1(q^2)
}{1-q^2/(m_B+m_X)^2} =
\frac{m_B+m_X}{2\,\sqrt{m_B\,m_X}} h_{A_1}(y(q^2))
$$
In the exact symmetry limit, $R_1=R_2=1$, $h_{A_1}=\iw$.
For the broken symmetry $R_1$, $R_2$ develop mild $y$ dependence
and their values at $y=1$ deviate from unity. The latter is
a stronger effect. Also $h_{A_1}$ may differ slightly from the
Isgur-Wise function $\iw$.

CLEO-II investigates full four-dimensional
correlations in all kinematical variables
describing the $B\to D^* \lem\bar{\nu}_l$ decay.~\cite{CLEOhqet}\
Since the data statistics are limited, $R_1$ and $R_2$ are
assumed to be constant and a linear $y$ dependence is assumed for
$h_{A_1}$ ($\approx 1-\rho^2\,(y-1)$).
As show in Fig.~\ref{fig:CLEOhqet}\ an excellent fit to the data
is obtained. Thus, we conclude that HQET works at the present
level of experimental sensitivity.
Numerical values obtained for the fitted parameters,
$R_1=1.18\pm0.30\pm0.12$,
$R_2=0.71\pm0.22\pm0.07$,
are in good agreement with the predictions based on
broken heavy quark symmetry:
$R_1=1.15-1.35$, $R_2=0.79-0.91$.~\cite{NeuClose}\
Also the value of the
slope parameter, $\rho^2=0.91\pm0.15\pm0.06$,
is consistent with
the HQET calculations and the
results of the fits to the $y$ variable in the
$|\Vcb|$ determinations.~\cite{Neubert}

\subsection{Semi-exclusive $B\to D^{**} \lem\bar{\nu}_l$.}

Comparison of the measured inclusive semileptonic branching ratio
with the exclusively measured branching ratios in decays to $D$
and  $D^*$
reveals that about $(35\pm7)\%$ of all semileptonic decays must
produce higher excitations of the charm mesons, and/or non-resonant
states. We have been giving them a generic label of  decays to
\lq\lq $D^{**}$'' states.

In addition to the two $S-$wave states ($D-1^1S_0$, $D^*-1^3S_1$),
four $P-$wave states are expected ($1^1P_1$, $1^3P_0$, $1^3P_1$,
$1^3P_2$).
Experimentally existence of
two narrow $P$ states has been established:
$D_1(2420)$ (decays to $D^*\pi$)
and $D_2^*(2460)$ (decays to $D^*\pi$ and $D\pi$).
HQET predicts that the other two states are wide and decay to
$D\pi$ and $D^*\pi$ respectively.

\begin{figure}[tbhp]
\vbox{
\vskip-0.1cm\noindent
(a) OPAL
\vskip-22.0cm\noindent
\hbox{\hskip-4.3cm
\psfig{file=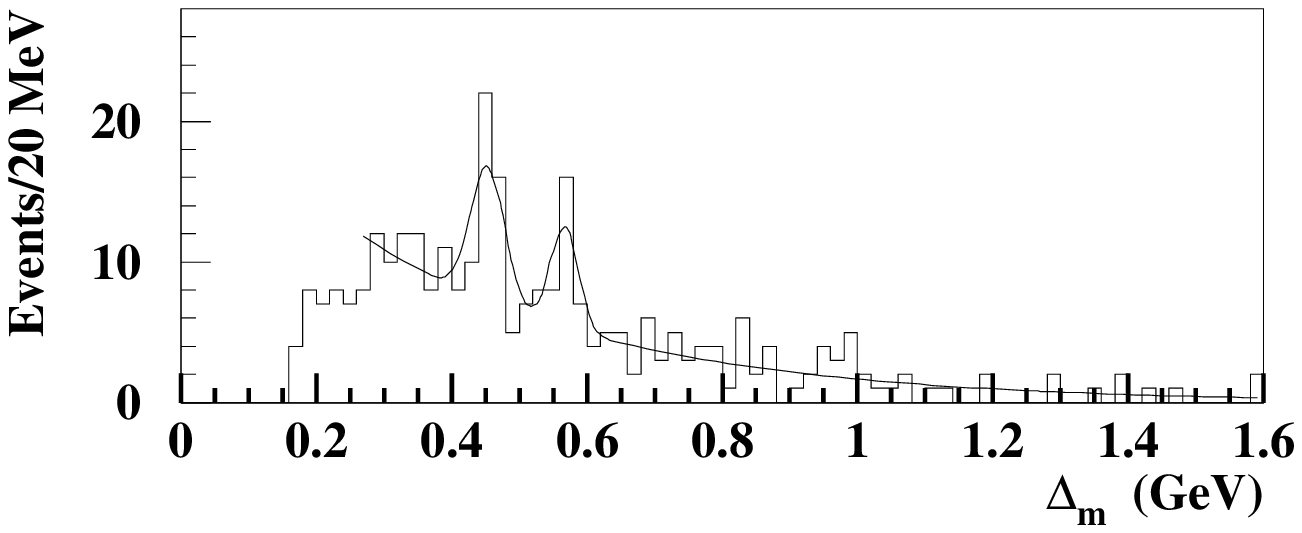,width=24.1cm,height=35.7cm}}
\vskip-8cm\noindent
(b) ALEPH
\vskip-13.3cm\noindent
\hbox{\hskip-0.3cm
\psfig{file=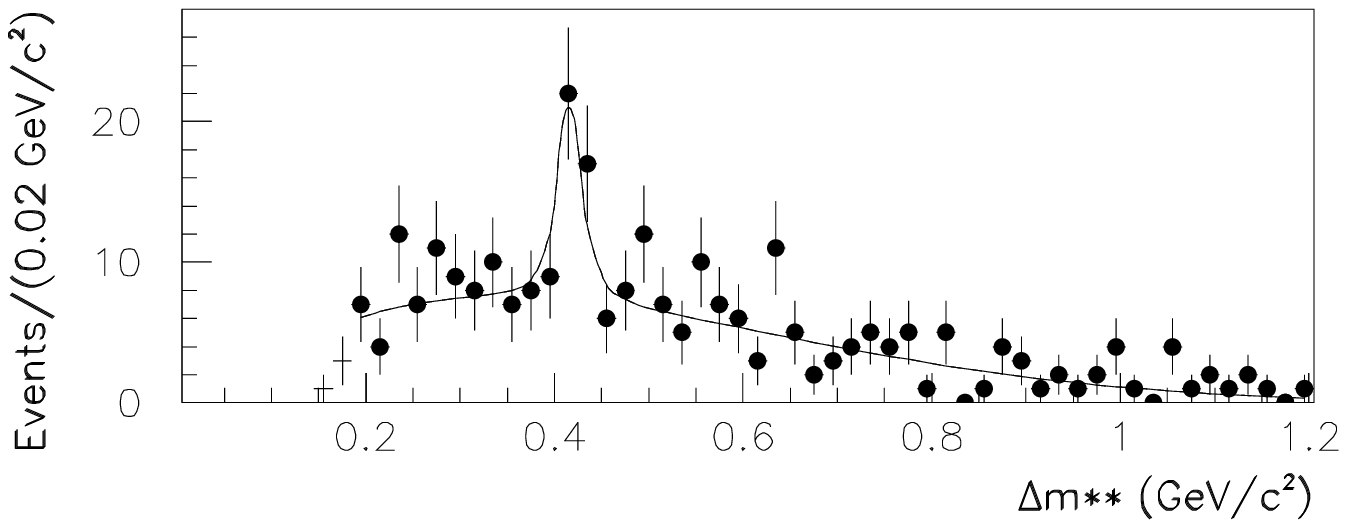,width=17.3cm,height=34.3cm}}
\vskip-15.4cm\quad}
\fcaption{
Mass difference $M({D^0\pi^+})-M({D^0})$
as observed for $B\to l\bar{\nu}_l D^+_JX$, $D^+_J\to D^0\pi^+$
candidates by OPAL (a) and ALEPH (b).
The first peak in the OPAL and ALEPH data is interpreted as
a sum of the $D^+_1$ and $D^{*+}_2$ states decaying to
$D^{*0}$ which in turn always decays to $D^0\gamma$ or $D^0\pi^0$
($\gamma$ and $\pi^0$ are not detected).
OPAL measures
$\BR(B\to l\bar{\nu}_l D^+_JX)\,\BR(D^+_J\to
D^{*0}\pi^+)=(1.8\pm0.6)\%$,
whereas ALEPH obtains a smaller value, $(0.5\pm0.2\pm0.1)\%$.
OPAL interprets the second peak in their data as evidence
for $D_2^{*+}$ production followed by the $D_2^{*+}\to D^0\pi^+$ decay
with a product branching ratio of
$\BR(B\to l\bar{\nu}_l D^{*+}_2 X)\,\BR(D^{*+}_2\to
D^{0}\pi^+)=(1.1\pm0.3^{+0.2}_{-0.3})\%$.
ALEPH finds no evidence for such decays and sets an upper limit
of $<0.2\%$ at 95\%\ C.L.
The branching ratios given here assume $f_{B^0}=f_{B^-}=0.4$ in $Z^0$ decays.
}
\label{fig:dss}
\vskip-0.5cm\quad
\end{figure}

First evidence for semileptonic $B$ decays to the $D^0_1$ state
was obtained by the ARGUS experiment.~\cite{ARGUSds}
Ability to correlate soft charged pions with the $B$ decay vertex
(same as the $D^{**}$ decay vertex) is a powerful experimental tool
available in the $Z^0$ data.
There is a wealth of new information from the LEP experiments.
DELPHI~\cite{DELPHIdss}\ measured decay rate to $D^0_1$ and $D_2^{*0}$.
ALEPH observes semileptonic decays to the $D_1$ state
in both charge modes.~\cite{ALEPHdss}
OPAL~\cite{OPALdss}\ claims the $D_2^*$ signal in both charged modes,
but the production of the $D_2^{*+}$ state is not confirmed by ALEPH
as illustrated in Fig.~\ref{fig:dss}.
CLEO-II~\cite{CLEOdss}\
sets only upper limits on $D^0_1$ and $D_2^{*0}$ production
but their results are not inconsistent with the LEP measurements.
ALEPH also presents evidence for production of wide resonances or
non-resonant states in $D^{*+}\pi^-$ and $D^0\pi^+$.

\newpage
All of these searches are semi-exclusive i.e. production of
other particles in addition to the observed states in the same
semileptonic decays is not excluded by the experiments.
Consequently the charge of the parent $B$ meson is not determined.
Detection efficiency calculated by the experiments depends
on the assumptions made about the production mechanism.
Furthermore, $D_1$ and $D_2^*$ decay branching fractions are
not known experimentally. All these factors together make
interpretation of the experimental results very difficult.
We conclude that the experiments confirm production of charm states
beyond the $D$ and $D^*$ mesons in semileptonic $B$ decays, but we are
still far from quantitative understanding of exclusive branching
fractions in this sector.

\section{Summary of $|\Vcb|$ determinations.}

\begin{figure}[hbtp]
\vbox{
\quad\vskip-2.3cm
\hbox{\hskip1cm\psfig{file=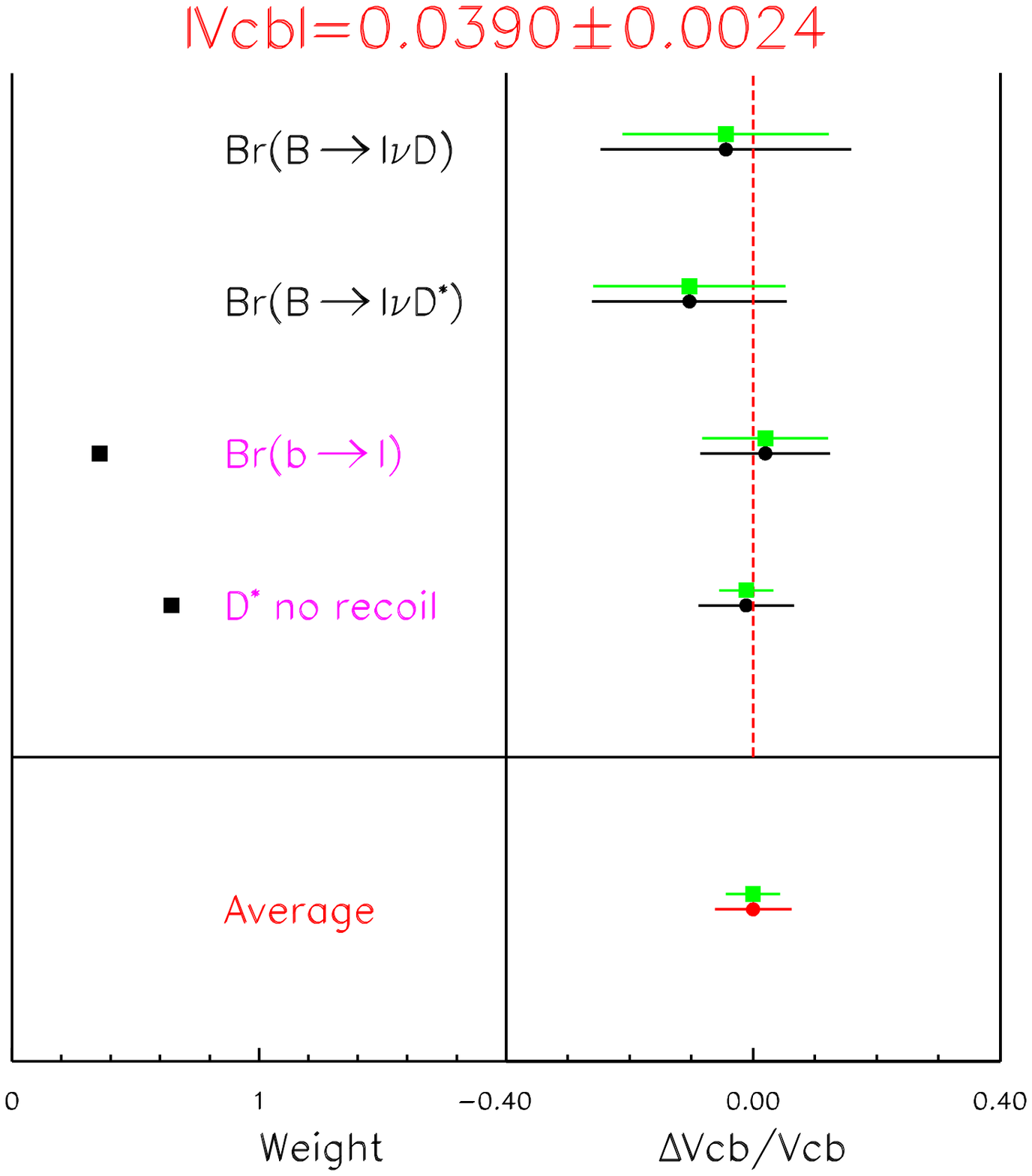,width=12.5cm,height=12.5cm}}
\vskip-2.3cm\quad}
\fcaption{Comparison of various determinations of $|\Vcb|$.
Relative deviations of each method from the average value are
shown.
The top errors show theoretical uncertainty alone.
The bottom errors show experimental and theoretical
errors combined.
The determinations from exclusive branching ratios are not used
in calculating the average since
the theoretical uncertainties
in these determinations
have not been systematically evaluated.
Weights in which the other two methods contribute to the average are
also shown.
}
\label{fig:vcb}
\vskip-0.5cm\quad
\end{figure}

Fig.~\ref{fig:vcb} summarizes all measurements of $|\Vcb|$ discussed
above. The determinations from the inclusive semileptonic
rate and from the
$D^*$ production rate measured at the zero recoil point
agree remarkably well.
Even though the determinations from the exclusive
semileptonic branching ratios are also consistent,
we average only the former two results
for the best estimate of $|\Vcb|$, since
the theoretical uncertainties in predictions
of the exclusive branching fractions
have not been systematically evaluated;
$$ |\Vcb|=0.0390^{+0.0016}_{-0.0014}\pm0.0017 $$
where the first error is experimental and the second theoretical.

\section{Determination of $|\Vub|$.}

\subsection{Inclusive semileptonic decays,
           $b\to u \lem\bar{\nu}_l$.}

The rate for
inclusive semileptonic decays of the $b$ quark to the $u$ quark
is by two orders of magnitude smaller than for decays to
the $c$ quark.
Inclusive $b\to u \lem\bar{\nu}_l$ decays are very difficult to
observe except for the endpoint of the lepton momentum
spectrum that extends above the kinematic limit in
the $b\to c \lem\bar{\nu}_l$ decays (due to $m_u\ll m_c$).
These decays were first observed
by CLEO-I.5~\cite{CLEObuold}\ and confirmed by ARGUS~\cite{ARGUSbu},
and later they were measured with a higher
precision by CLEO-II.~\cite{CLEOIIbu}\
Unfortunately,
determination of $|\Vub|$ from these
measurements suffers from a large extrapolation factor
from the endpoint rate to a total branching ratio
that is not well predicted theoretically:~\cite{CLEOIIbu}
$$ \frac{|\Vub|}{|\Vcb|}=0.08\pm0.01\pm0.02 $$
The first error is experimental, the second reflects the
model dependence in the extrapolation factor.
Model dependence is large because the form factors
in $b\to u$ decays must
be predicted over a much larger range of $q^2$ than
in the $b\to c$ decays.
Different models disagree even on the relative
contributions of various exclusive modes to the endpoint
region;
the predicted ratio of rates for  $B\to \rho \lem\bar{\nu}_l$ and
$B\to \pi \lem\bar{\nu}_l$ decays varies from
1.5~\cite{ISGW2}\ to 4.6.~\cite{KS}\
It has been also suggested that
non-resonant $\pi\pi$ production could
significantly contribute to the endpoint region.~\cite{ramirez}
Experimental information on dynamics of  $b\to u \lem\bar{\nu}_l$
decays is needed to make progress in  determining $|\Vub|$.

\subsection{Exclusive semileptonic decays,
           $B\to (\pi,\rho,\omega) \lem\bar{\nu}_l$.}

\begin{figure}[tbhp]
\vbox{\vskip-1cm \noindent
\phantom{a}
\hskip2cm (a)\quad $B\to \lem\bar{\nu}\pi$ \hskip5cm
(b)\quad $B\to \lem\bar{\nu}\rho,\omega$
\vskip-9.6cm\noindent
\hbox{
\psfig{file=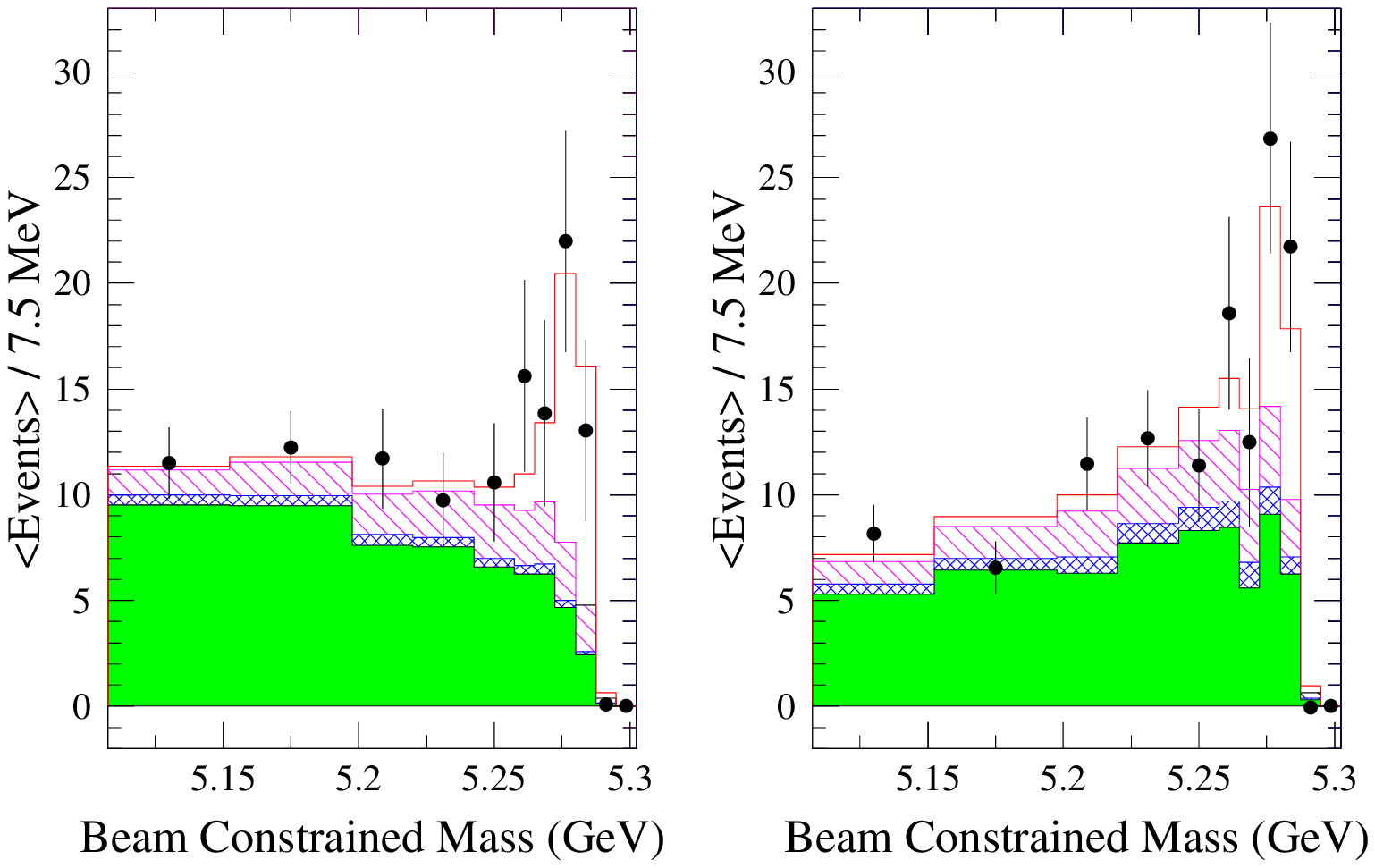,width=20cm,height=15cm}
}
\par\noindent
\quad \hskip2cm  (c)\quad $B\to \lem\bar{\nu}\rho$ \hskip5cm
    (d)\quad $B\to \lem\bar{\nu}\pi^0\pi^0$
\vskip-9.6cm\noindent
\hbox{
\psfig{file=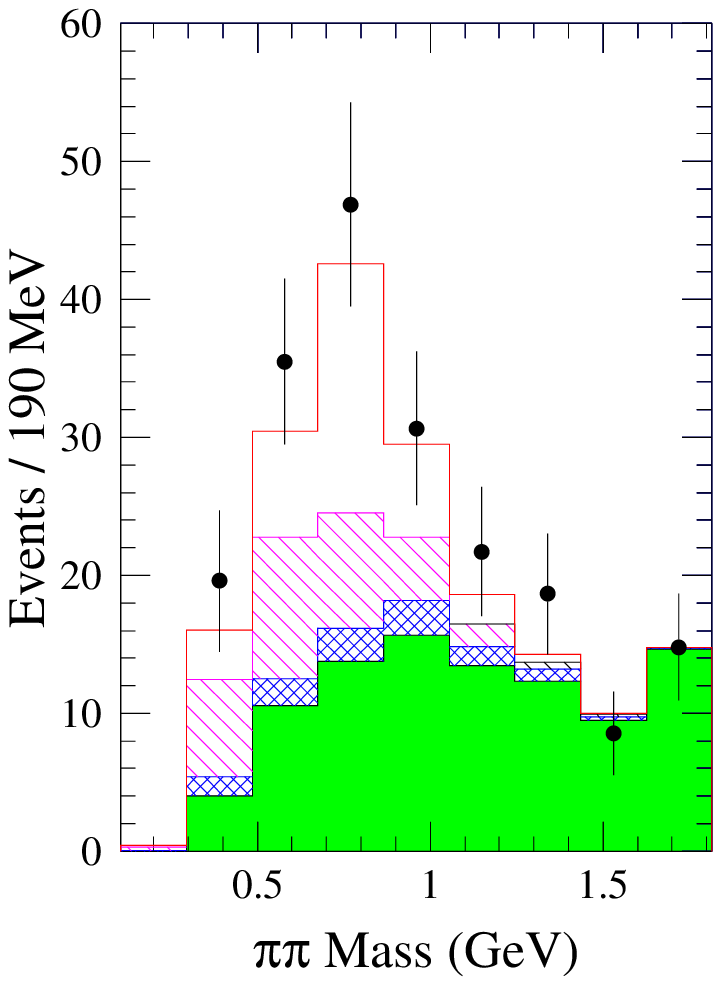,width=20cm,height=15cm}
\hskip-12.76cm
\psfig{file=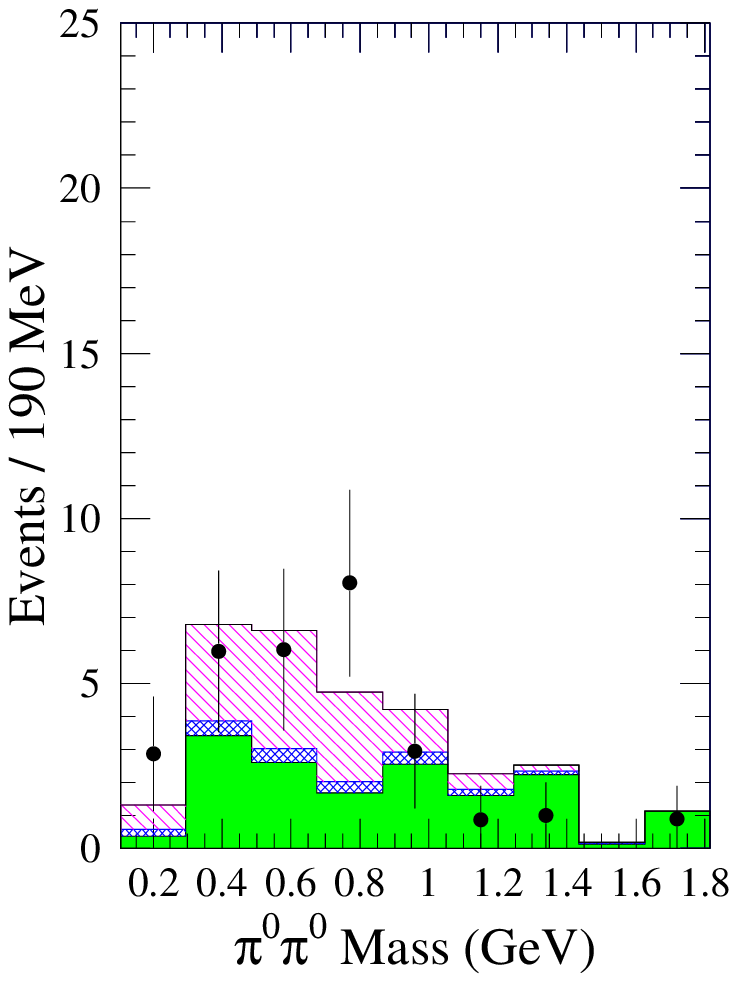,width=20cm,height=15cm}
}
\par\noindent
\phantom{a}\hskip2cm  (e)\quad $B\to \lem\bar{\nu}\pi$
   \hskip5cm  (f)\quad $B\to \lem\bar{\nu}\rho,\omega$
\vskip-9.6cm\noindent
\hbox{
\psfig{file=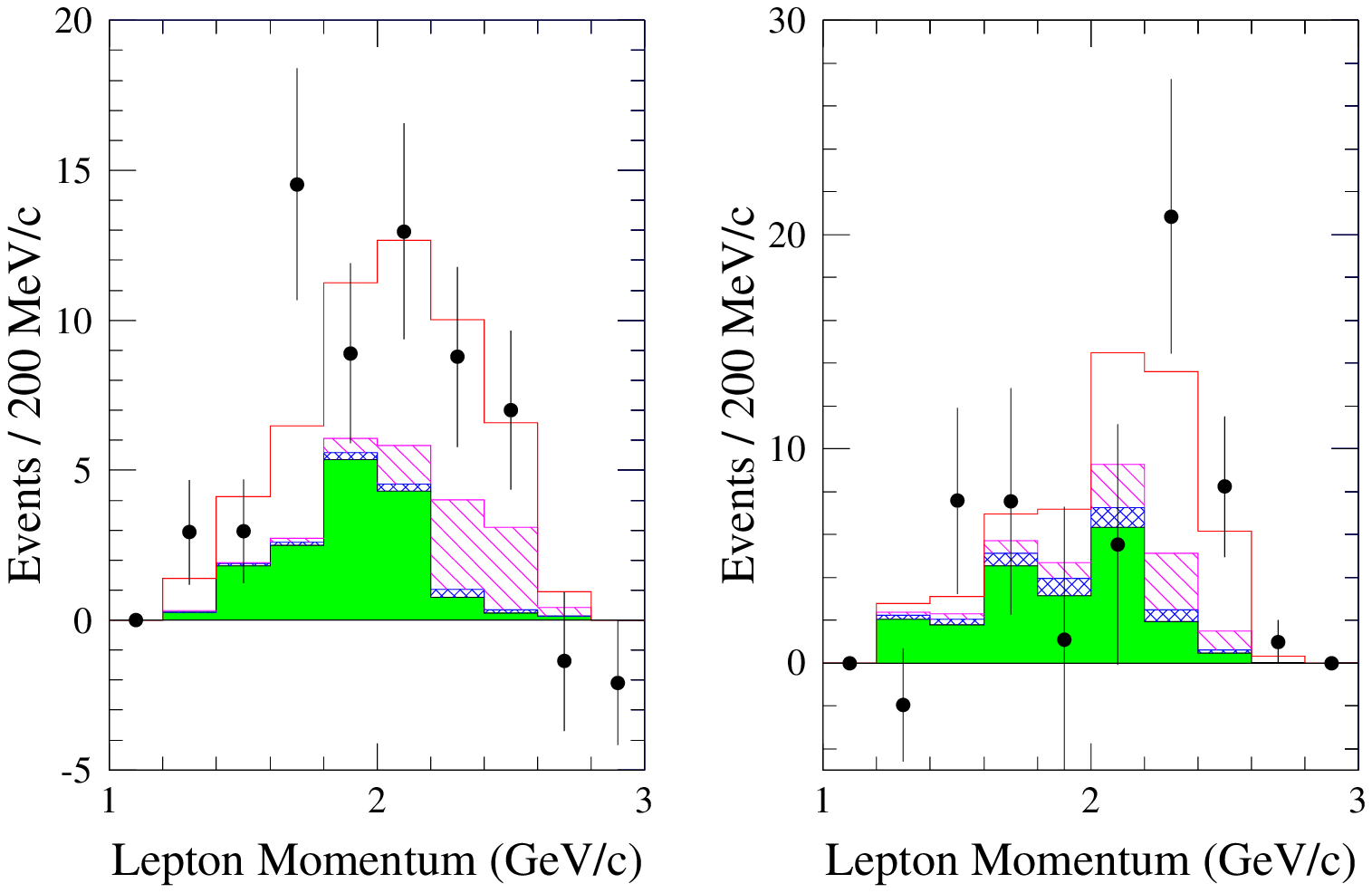,width=20cm,height=15cm}
}
\vskip-0.5cm\quad}
\fcaption{
The data and the projections of the fit in the CLEO-II
analysis of exclusive $B\to (\pi,\rho,\omega)\lem\bar{\nu}_l$ decays.
The points are continuum subtracted data.
The histograms show the
contributions from
$b\to c \lem\bar{\nu}_l$ (shaded),
feed-down from higher mass $b\to u \lem\bar{\nu}_l$ (cross-hatched),
signal mode cross-feed (hatched) and signal (hollow).
(a),(b) Significant peaks are observed at the $B$ meson mass.
(c),(d) Decays to the $\rho$ resonance are observed with no
        evidence for non-resonant production of $\pi\pi$ states.
(e),(f) As expected, lepton momentum spectrum is harder in decays
        to the vector states.}
\label{fig:exclbu}
\vskip-0.5cm\quad
\end{figure}
CLEO-II presents~\cite{CLEOIIbuexcl}\
the first observation of exclusive semileptonic
decays  $B\to (\pi,\rho,\omega) \lem\bar{\nu}_l$.
Previously, the most sensitive search for these decays reported
only upper limits.~\cite{CLEOIIbuexclold}
Suppression of  $b\to c \lem\bar{\nu}_l$ and continuum backgrounds
has been improved by reconstruction of the neutrino four-vector
from missing energy and momentum in the event.
This method has been already outlined in Sec.~\ref{sec:d0lnu}.
Thanks to the smaller $b\to c$ background,
the lepton momentum range is extended below the endpoint
region. As in any reconstruction of exclusive $B$ decays
at $\Upsilon(4S)$, the beam energy constraint and the beam constrained
$B$ mass are essential tools in signal extraction.
For signal events,
$\delta E=E_{beam}-(E_{X_u}+E_l+|\vec{p}_\nu|)$ peaks at zero, and
$M_B=\sqrt{E_{beam}^2-(\vec{p}_{X_u}+\vec{p}_l+\vec{p}_\nu)^2}$ peaks
at the true $B$ meson mass ($X_u=\pi$, $\rho$, or $\omega$).
Background events populate a wide range in these variables.
Two dimensional distributions, $\Delta E$ vs $M_B$,
are fit to unfold the signal from the backgrounds.
Because the signal branching ratios are small and the neutrino
reconstruction technique is costly in detection
efficiency (efficiency is a few percent per decay mode)
CLEO-II groups the signal channels into pseudo-scalar and vector
modes using isospin symmetry:
$\frac{1}{2} \Gamma(\bar{B}^0\to \pi^+\lem\bar{\nu})  =
            \Gamma({B}^-\to \pi^0\lem\bar{\nu})$,
$\frac{1}{2} \Gamma(\bar{B}^0\to \rho^+\lem\bar{\nu}) =
            \Gamma({B}^-\to \rho^0\lem\bar{\nu}) \approx
            \Gamma({B}^-\to \omega\lem\bar{\nu})$.
All five decay modes are fit simultaneously.
The signal shape in the $\Delta E$, $M_B$ plane is taken from the Monte
Carlo simulation.
Thus, there are two free parameters in the fit related to the
signal branching fractions.
The largest background comes from  $b\to c \lem\bar{\nu}_l$
decays (e.g. with undetected $K^0_L$ or second $\nu$).
The shape of this background is fixed from the Monte Carlo
simulation and the rate is a free parameter in the fit
(alternatively five free parameters are used allowing different
scale factors in each mode).
Continuum background is small and is subtracted using the
data taken below the $\Upsilon(4S)$ resonance.
Feed-across between the signal modes is large and is predicted
by the Monte Carlo with the scale fixed by the signal parameters.
Feed-down from higher mass $b\to u \lem\bar{\nu}_l$ decay modes
is small and it is estimated from the inclusive endpoint
analysis with help of the Monte Carlo.
Projections of this fit onto several variables of interest are
shown in Fig.~\ref{fig:exclbu}.
Significant signals are observed for both pseudo-scalar
and vector decay modes (3.9 and 6.4 standard deviations respectively).
Presence of the $\rho$ resonance is evident (Fig.~\ref{fig:exclbu}c)
and the results for $\omega$ are consistent with the scaling from
the $\rho$ channels.
No excess of events over the background estimates
is seen in the
$\pi^0\pi^0$ channel that would have been evidence
for non-resonant production (Fig.~\ref{fig:exclbu}d).
Since the detection efficiency is model dependent, CLEO-II
presents branching fraction results for
various models.
The measured branching fractions are consistent with
the model predictions and the
$|\Vub|$ extracted from the inclusive
data as shown in Table~\ref{tab:bu}.
Both the experimental errors and the model dependence are large.
Nevertheless, this measurement constitutes an important milestone
in the quest for better determination of $|V_{ub}|$.
With increased experimental statistics
exclusive reconstructions of semileptonic $b\to u$ decays will
shed light onto strong dynamics of these decays, and consequently
help develop more reliable models of inclusive decays.
\begin{table}[bthp]
\tcaption{Branching ratios for exclusive $b\to u \lem\bar{\nu}_l$ decays
measured by CLEO-II compared to the model predictions.
The uncertainty
on $Br(B^0\to \rho^-\lep\nu)$
due to possible non-resonant background,
$^{+0}_{-20}\%$,  is not included below.
The model predictions are normalized by assuming $|\Vub|/|\Vcb|=0.08$
as indicated by the inclusive measurement.
The experimental branching ratios are model dependent via
the reconstruction efficiency.}
\label{tab:bu}
\small
     \def\1#1{\multicolumn{2}{c|}{#1}}
     \def\2#1{\multicolumn{2}{c||}{#1}}
     \def\meas{measured}
     \def\pred{predicted}
     \begin{tabular}{||l|ll|ll|ll||}
\hline\hline
   {} & \1{} & \1{} & \2{} \\
   Model
    &  \1{$Br(B^0\to \pi^-\lep\nu)$}  & \1{$Br(B^0\to \rho^-\lep\nu)$}  &
   \2{$Br(\rho^-)/Br(\pi^-)$} \\
   \cline{2-7}
   & \meas   & \pred
   & \meas   & \pred
   & \meas   & \pred  \\
{} & $10^{-4}$ & $10^{-4}$ & $10^{-4}$ & $10^{-4}$ & {} & {} \\
\hline
{} & {} & {} & {} & {} & {} & {} \\
   ISGW\cite{ISGW}
              &  $1.3\pm0.4$         & $0.3$
              &  $2.3\pm0.7$         & $1.3$
              &  $1.7^{+1.0}_{-0.8}$ & $4.0$
              \\
   ISGW2\cite{ISGW2}
              &  $2.0\pm0.7$         & $1.5$
              &  $3.2\pm0.9$         & $2.2$
              &  $1.6^{+0.9}_{-0.7}$ & $1.5$
              \\
   KS\cite{KS}
              &  $1.6\pm0.5$         & $1.1$
              &  $2.7\pm0.8$         & $5.1$
              &  $1.7^{+1.0}_{-0.7}$ & $4.6$
              \\
   WSB\cite{WSB}
              &  $1.6\pm0.6$          & $1.0-1.6$
              &  $3.9\pm1.1$          & $2.9-6.6$
              &  $2.4^{+1.4}_{-1.1}$  & $3.0-4.3$  \\
{} & {} & {} & {} & {} & {} & {} \\
      \hline\hline
\end{tabular}
\end{table}

\section{Conclusions}

Decays of the $b$ quark play a unique role for physics of the
Cabibbo-Kobayashi-Maskawa matrix.
This is also an important place to test our understanding
of interplay between strong
and weak interactions.
Higher order decays of the $b$ quark are also a sensitive window
for physics beyond the Standard Model.

Studies of the most common semileptonic
$b\to c$ decays is a mature field both
theoretically and experimentally. Measurements of $B\to D^*
\lem\bar{\nu}_l$ rate at zero recoil point by CLEO-II, together with
two new measurements by ALEPH and DELPHI at LEP dominate the world
average value of $|\Vcb|$. Even though $|\Vcb|$ is one of the better
known elements of the CKM matrix, impact on the
matrix unitarity tests motivates the push for even better
determinations of $|\Vcb|$ in the future.

The process of understanding
strong dynamics in semileptonic $b\to u$
decays have just started with the first observation of the exclusive
decay modes by CLEO-II. Eventually this  will lead
to a better determination of $|\Vub|$.

Higher order $b\to t\to s$ decays have been observed by CLEO-II
via $b\to s\gamma$ decays.
This establishes existence of processes mediated by the penguin
diagrams in an unambiguous way.
These measurements also
provided first direct results for $|\Vts|$ and plentiful
constraints on extensions of the Standard Model.
In addition to increased experimental statistics,
next-to-leading order theoretical calculations are needed for
better constraints in the future.
Siblings of the electromagnetic penguin, gluonic and electroweak
penguins, should be experimentally established in the next few years.

Similar $b\to t\to d$ decays, that would provide a measurement
of $|\Vtd|$, have not been yet observed.
These decays will be in the detection range of the next generation of
$b$ experiments at $e^+e^-$ colliders already under construction.

Studies of $b$ decays, together with $B\bar{B}$ oscillations and
CP$-$violations, which are discussed in separate papers at this
conference, are a dynamic experimental field.
In addition to the upgraded CLEO experiment (CLEO-III),
two new asymmetric $e^+e^-$ colliders are under construction at
KEK and SLAC with the main goal of detecting the CP violation in $b$
decays. Enormous exploratory potential also exists at machines with
high energy hadronic beams. HERA-B experiment is under construction
at DESY. Hopefully Tevatron and LHC will also exercise in full
their reach in $b$ quark physics.

\section{Acknowledgements}

I would like to thank the people who provided me with
the information on recent experimental results,
in particular:
Chafik Benchouk, Concezio Bozzi, Roger Forty, Pascal Perret  and
Vivek Sharma from ALEPH,
Ronald Waldi from ARGUS,
Fritz Dejongh from CDF,
Bob Clare and David Stickland from L3,
Wilbur Venus from DELPHI,
Andrzej Zieminski from D0,
Chris Hawkes and Bob Kowalewski from OPAL,
Marina Artuso, Ken Bloom, Lawrence Gibbons,  Klaus Honscheid,
Chris Jones, Suzanne Jones,
Scott Menary, Giancarlo Moneti, Hitoshi Yamamoto
and the other colleagues from CLEO.
I would also like to thank
Matthias Neubert and Joe Kroll for useful exchange of information.
Last but not least, I would like to thank Sheldon Stone for
valuable suggestions concerning this article.

\section{References}
\def\etal{{\it et al.}}
\def\eps#1{Paper contributed to {\it this conference} and
      to the {\it EPS Conference}, {\bf Brussels} (1995) No.#1}
\def\glasgow#1{Paper contributed to {\it the XXVII Int.
      Conf. on High Energy Phys.}
      {\bf Glasgow},
      (1994) No.#1}
\def\ithaca{Paper contributed to {\it the
      XVI International Symposium on Lepton
      and Photon Interactions}, {\bf Ithaca}, NY,
      (1993)}
\def\ppr{Preprint}

\end{document}